\documentclass[a4paper,11pt]{article}
\pdfoutput=1 
\usepackage{jcappub} 
\usepackage[T1]{fontenc} 
\usepackage{siunitx}
\usepackage{lineno}
\usepackage{longtable}
\usepackage{lscape}
\usepackage{graphicx}
\usepackage{threeparttable}
\usepackage{booktabs}
\usepackage{afterpage}

\title{\boldmath JUNO sensitivity to the annihilation of MeV dark matter in the galactic halo}
\collaborationImg{\includegraphics[width=2cm,height=2cm]{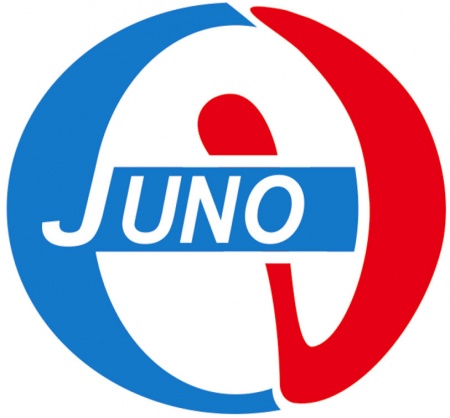}}
\collaboration{The JUNO collaboration}
\emailAdd{Juno\_pub\_comm@juno.ihep.ac.cn}
\author[5,41]{Angel Abusleme}
\author[46]{Thomas Adam}
\author[67]{Shakeel Ahmad}
\author[67]{Rizwan Ahmed}
\author[56]{Sebastiano Aiello}
\author[67]{Muhammad Akram}
\author[67]{Abid Aleem}
\author[49]{Tsagkarakis Alexandros}
\author[29]{Fengpeng An}
\author[22]{Qi An}
% ...
\affiliation[5]{Pontificia Universidad Cat\'{o}lica de Chile, Santiago, Chile}
% ...
\affiliation[22]{University of Science and Technology of China, Hefei, China}
% ...
\affiliation[29]{East China University of Science and Technology, Shanghai, China}
% ...
\affiliation[41]{Millennium Institute for SubAtomic Physics at the High-energy Frontier (SAPHIR), ICN2019\_044, ANID, Chile, cl}
% ...
\affiliation[46]{IPHC, Universit\'{e} de Strasbourg, CNRS/IN2P3, F-67037 Strasbourg, France}
% ...
\affiliation[49]{III. Physikalisches Institut B, RWTH Aachen University, Aachen, Germany}
% ...
\affiliation[56]{INFN Catania and Dipartimento di Fisica e Astronomia dell Universit\`{a} di Catania, Catania, Italy}
% ...
\affiliation[67]{Pakistan Institute of Nuclear Science and Technology, Islamabad, Pakistan}
% ...

\abstract{We discuss JUNO sensitivity to the annihilation of MeV dark matter in the galactic halo via detecting inverse beta decay reactions of electron anti-neutrinos resulting from the annihilation. 
We study possible backgrounds to the signature, including the reactor neutrinos, diffuse supernova neutrino background, charged- and neutral-current interactions of atmospheric neutrinos, backgrounds from muon-induced fast neutrons and cosmogenic isotopes. 
A fiducial volume cut, as well as the pulse shape discrimination and the muon veto are applied to suppress the above backgrounds. It is shown that JUNO sensitivity to the thermally averaged dark matter annihilation rate in 10 years of exposure would be significantly better than the present-day best limit set by Super-Kamiokande and would be comparable to that expected by Hyper-Kamiokande. }
\begin{document}
\maketitle
\flushbottom
%----------- Section 1 Introduction ----------
\section{Introduction}
\label{sec:intro}
The existence of non-baryonic Dark Matter (DM) in the Universe has been well established by astronomical observations. For most spiral galaxies, the rotation curve of stars or gases far from the galactic center does not decline with increasing distance but rather stays as a constant. This strongly indicates the existence of a massive dark halo which contains the galactic disk and extends well beyond the size of the visible part of the galaxy~\cite{Rubin:1970}.
One promising DM candidate is the Weakly Interacting Massive Particle (WIMP)~\cite{Jungman:1995df}, which predicts a correct relic density based on the weak interaction annihilation cross section. In the WIMP scenario, the thermally averaged self-annihilation rate $\langle\sigma v\rangle$ of DM is predicted to be \num{3e-26}\si{\cubic\centi\meter\per\second} regardless of the annihilation channel. 

Quarks and leptons could be produced through DM annihilation and therefore provide foundations for numerous DM indirect search experiments~\cite{Super-Kamiokande:2004pou,IceCube_DMind,Fermi,AMS2,HESS}. 
In this paper, we focus on the scenario in which neutrinos are produced directly from DM  annihilation, $\chi\chi\to \nu\bar{\nu}$. For DM candidates lighter than the muon, this is the only neutrino production channel. For heavier DM candidates, neutrinos can also arise from decays of hadronic or leptonic final states produced by the annihilation.      

Searches for neutrinos originating from DM annihilation have been proposed for astrophysical neutrino observatories~\cite{NuDM_MeV}, accelerator neutrino experiments~\cite{Accer,Coiilders} and solar neutrino measurements~\cite{SolarNu}.
Several experiments including KamLAND~\cite{KamLAND,KamLAND_2022}, Super-Kamiokande (SuperK)~\cite{SKColab_DM}, IceCube~\cite{IceCUBEDM_Co} and ANTARES~\cite{IceCUBE_DMEXP} have searched for neutrino signatures from DM annihilation. Among them, KamLAND~\cite{KamLAND_2022} obtained the updated $90\%$ confidence level upper limit $\langle\sigma v\rangle=$ \SIrange[range-units = brackets, range-phrase=--]{1}{11}{}$\times$$\num{e-26}\si{\cubic\centi\meter\per\second}$ for DM mass in the range of \SIrange[range-units = brackets, range-phrase=--]{9}{21}{MeV} with a nominal angular-averaged intensity of the galactic DM profile. More stringent limits on $\langle\sigma v\rangle$ were deduced from the data of SuperK~\cite{SuperK_782, SuperK_1542}, while IceCube high-energy cosmic neutrino data can be used to perform the imaging of galactic  DM~\cite{Nu_Glalt_Icecube}. Finally, the expected sensitivity to $\langle \sigma v\rangle$ is also discussed for Hyper- Kamiokande (HyperK)~\cite{HyperK}. 

JUNO~\cite{JUNO:2015zny,Muon_strategies}, equipped with a central acrylic sphere containing \SI{20}{\kilo\tonne} of liquid scintillator (LS), will significantly improve the sensitivity to $\langle\sigma v\rangle$.
In this paper, we discuss the JUNO sensitivity to the neutrino flux from DM annihilation $\chi\chi\to \nu\bar{\nu}$ in the galactic halo~\cite{Palomares-Ruiz:2007trf} where final state neutrinos are monochromatic i.e. $E_{\nu}$ = $m_{\chi}$. We specifically focus on the DM mass range of \SIrange[range-units = brackets, range-phrase=--]{15}{100}{MeV} where the lower mass limit is set to avoid the reactor neutrino background while the upper mass limit allows us to consider only the direct annihilation channel for neutrino productions. We note that JUNO is also capable of measuring more energetic neutrino events, such as those arising from heavier DMs through more complicated neutrino production mechanisms or atmospheric neutrino events in the energy range between $0.1$ GeV and $10$ GeV~\cite{Atmos_mu_e}.

We shall first evaluate the inverse beta decay (IBD) signature in JUNO arising from DM annihilation $\chi\chi\to \nu\bar{\nu}$ in the galactic halo. Backgrounds dominating or comparable to the signal will be investigated hereafter. Among them, the events induced by atmospheric neutrinos interacting with $^{12}$C nuclei through neutral-current (Atm-$\nu$ NC) are dominant. We perform pulse shape discrimination (PSD) to reject such backgrounds. Considering all backgrounds and the overall event selection efficiency, we find that the JUNO sensitivity to $\langle \sigma v\rangle$ can reach $\sim$ \num{1e-25}\si{\cubic\centi\meter\per\second} with a 10 years of exposure time, which will be competitive with any other existing or upcoming detectors in the near future. 

The paper is organized as follows: In Sec. 2 we give a brief introduction of the JUNO detector. In Sec. 3 we present predictions on the DM annihilation rate in the galactic halo and the IBD event rate in the JUNO detector. The background studies are presented in Sec. 4, with efficiencies of various veto methods evaluated. Sec. 5 focuses on PSD methods employed for separating IBD and non-IBD events. This is particularly useful for suppressing backgrounds from Atm-$\nu$ NC events mentioned before. In Sec. 6 we present the JUNO sensitivity to $\langle \sigma v\rangle$ for the DM mass range of \SIrange[range-units = brackets, range-phrase=--]{15}{100}{MeV}. We summarize and conclude in Sec. 7.
%----------- Section 2 JUNO ----------
\section{JUNO detector}
JUNO is a multi-purpose underground liquid scintillator (LS) detector that aims to decipher the neutrino mass ordering as the primary goal. The low muon rate, \SI{0.004}{\per\second\per\square\meter}, is due to the \SI{700}{\meter} rock overburden that shields the detector from the flux of cosmic muons with \SI{207}{\giga\electronvolt} of average energies, and therefore makes the detector suitable for searching exotic sources of neutrinos. 

\begin{figure}
\centering
\includegraphics[width=0.7\textwidth]{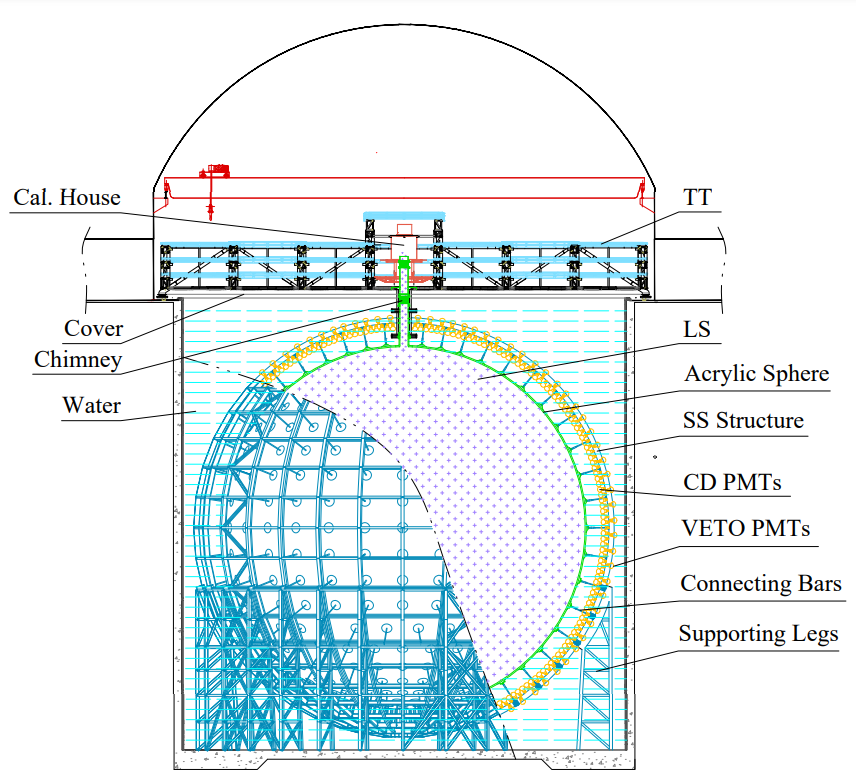}
\caption{Scheme of the JUNO detector}
\label{figure_detector_scheme}
\end{figure}

JUNO comprises central and veto detectors as illustrated in Fig.\ref{figure_detector_scheme}~\cite{Muon_strategies}. The central detector is a spherically shaped acrylic shell with a \SI{17.7}{\meter} radius filled with \SI{20}{\kilo\tonne} of liquid scintillator. There are 17,612 \SI[parse-numbers=false]{20}{\text{-inch}} and 25,600 \SI[parse-numbers=false]{3}{\text{-inch}} PMTs mounted outside the acrylic ball, which provide around $78\%$ photocathode coverage and excellent 3\%/$\sqrt{E({\rm MeV})}$ energy resolution~\cite{JUNO_Calibration}.
The veto detector is composed of a top tracker and a water Cherenkov detector.  The top tracker~\cite{JUNO_Top_Tracker} covers half of the top surface of the water pool as shown in figure ~\ref{figure_detector_scheme}. It is composed of 3-layer plastic scintillators originally used in the OPERA~\cite{OPERA:2000zdf,OPERA_TOP_TRACKER} experiment and re-utilized by JUNO. The water Cherenkov detector is a cylindrical water pool with \SI{43.5}{\meter} in both diameter and height. It is filled with \SI{30}{\kilo\tonne} of ultrapure water and maintained by a circulation system. There are 2400 \SI[parse-numbers=false]{20}{\text{-inch}} PMTs mounted on the stainless steel frame to detect the Cherenkov light of cosmic muons. Together with veto and central detectors, the tracks of cosmic muons can be precisely reconstructed and therefore a high efficiency (99.5\%) on muon tagging is achieved. The low muon rate and high muon tagging efficiency enable the detector to perform rare event searches.
%----------- Section 3 DM ----------
\section{Neutrino signature from DM annihilation in the Milky Way}
The estimation of the DM-induced neutrino event rate is based on two assumptions. First, we assume that DM annihilates $100\%$ into neutrino-antineutrino pairs, which sets an upper limit for the neutrino flux resulting from DM annihilation in the galactic halo. Second, we only consider the DM mass range of \SIrange[range-units = brackets, range-phrase=--]{15}{100}{MeV} which meets the selection criteria we set for DM search in JUNO.
The entire galactic halo is chosen as the target for maximizing the neutrino flux and at the same time minimizing the impact of profile uncertainties around the galactic center (GC).  

The $\bar{\nu}_e$ neutrino flux spectrum resulting from DM annihilation is given by~\cite{Palomares-Ruiz:2007trf}
\begin{equation}
\frac {\text{d}\phi^{\rm DM}_{\bar{\nu}_e}}{\text{d}E_{\bar{\nu}_e}} = 
\frac {1}{2} \langle\sigma v\rangle J_{\rm avg} \frac  {R_{\rm sc}\rho^2_0}{m^2_{\chi}} \frac {1}{3} \delta(E_\nu-m_{\rm \chi}),
\label{Neutrino_Flux}
\end{equation}
where $m_\chi$ is the DM mass, $R_{\rm sc}\sim$ \SI{8.5}{kpc} the distance between the GC and the solar system, and $\rho_0\equiv \rho(R_{\rm sc})$ the DM density in the local universe. The factor 1/2 comes from the assumption that DM is a Majorana particle while the factor 1/3 arises from the assumption that DM candidates annihilate into all three neutrino flavors with an equal probability.  The canonical value for the thermally averaged DM annihilation rate, $\langle\sigma v\rangle$, is \num{3e-26}\si{\cubic\centi\meter\per\second}. In the case that DM candidates annihilate predominantly into one particular flavor, neutrino flavor transitions occurring between the production and detection points would generate the other flavors with comparable intensities. Since we focus on detecting the annihilation channel $\chi + \chi \to \nu + \bar{\nu}$, the neutrino energy is equal to the DM mass and this is implemented by the delta function $\delta(E_\nu-m_{\chi})$.

The angular-averaged intensity, $J_{\rm avg}$, is an integration over the square of the DM density along the line of sight and normalized by the square of the local DM  density, 
\begin{equation}
J_{\rm avg}={{\frac {1}{2R_{\rm sc}{\rho_{0}}^{2}}}} 
\int_{-1}^{1} \textrm{d}\cos\psi \int_{0}^{\mathcal{\ell}_{\rm max}} \rho^2\left(r(l,\psi)\right)\textrm{d}\ell, 
\end{equation}
where $\rho$ is the DM density at the specific location described by the coordinate $(l,\psi)$ with $l$ the distance between the DM and the Earth while $\psi$ the direction of the DM viewed from the Earth with $\psi=0$ corresponding to the direction of GC. The distance between DM and GC is given by $r = \sqrt{R^{2}_{\rm sc} -2\ell R_{\rm sc} \cos\psi + \ell^2}$ while $\ell_{\rm max} =\sqrt{r_s^2 - \sin^2\psi R^2_{\rm sc}}+R_{\rm sc}\cos\psi$ with $r_s$ the radius of the galactic halo. The integration runs from $\psi$ = 0$^\circ$ to $\psi$ = 180$^\circ$ since it is challenging to precisely determine the neutrino direction with the JUNO detector at these energies.
In the galactic halo, one assumes a spherically symmetric DM density profile with isotropic velocity dispersion. Hence the DM profile can be parametrized as:
\begin{equation}
\rho(r)= \rho_0 {{\left(\frac {R_{\rm sc}}{r}\right)}^\gamma} 
{\left[\frac {1+{(R_{\rm sc}/r_{s})}^{\alpha}} {1+{(r/r_{s})}^{\alpha}}\right]}^{(\beta-\gamma)/\alpha}.
\end{equation} 
For the variable set ($\alpha$,$\beta$,$\gamma$,$r_s$), $\alpha$ determines the profile shape around $r_{s}$, $\beta$ is the slope in the limit $r\to \infty$, $\gamma$ is the inner cusp index, and $r_{s}$ is the halo radius. In Table~\ref{table:DMprofile_para}, we summarize the parameter values corresponding to three commonly used profiles, the Navarro-Frenk-White (NFW)~\cite{NFW1996}, Kravtsov~\cite{KKBP1998} and Moore~\cite{MQGSL1999} profiles. We adopt the benchmark case of \mbox{$J_{\rm avg}$ =  5.0~\cite{Palomares-Ruiz:2007trf}} for presenting our results.
\begin{table}
\centering
\caption[Parameters of NFW, MQGSL and KKBP dark matter halo profiles.]{Summary of parameters for different DM halo profiles.}
\label{table:DMprofile_para}
\begin{tabular}{ccccccc}

\hline
Halo profiles & $\alpha$ & $\beta$ & $\gamma$ & $r_s$ [kpc] & $\rho(R_{\text{sc}})$ $[{\rm GeV/cm^{3}}]$ & $J_{\rm avg}$ \\
\hline
NFW \cite{NFW1996}         & 1        & 3       & 1        & 20          & 0.3                         & 3  \\
MQGSL \cite{MQGSL1999}       & 1.5      & 3       & 1.5      & 28          & 0.27                        & 8  \\
KKBP  \cite{KKBP1998}       & 2        & 3       & 0.4      & 10          & 0.37                        & 2.6  \\
\hline
Canonical &    &  &  &  &  &  5.0~\cite{Palomares-Ruiz:2007trf} \\
\hline  \hline 
\end{tabular}
\end{table}

Finally, the number of neutrino events in JUNO is given by 
\begin{equation}
    \frac{\text{d}N_S(E_{\bar{\nu}_e})}{\text{d}E_{\bar{\nu}_e}} = \sigma_{\rm IBD}(E_{\bar{\nu}_e})\cdot\frac{\text{d}\phi_{\bar{\nu}_e}^{\rm DM}(E_{\bar{\nu}_e})}{\text{d}E_{\bar{\nu}_e}}\cdot N_{\rm target}\cdot t\cdot\epsilon~\ .
    \label{equation_theo_signal_spectrum}
\end{equation}
Here $\sigma_{\rm IBD}$ is the cross section for the IBD reaction, $\nu_e + p \rightarrow e^+ + n$, which leads to a prompt signal from positron-electron annihilation and a delayed signal from neutron capture. The value for $\sigma_{\rm IBD}(E_{\bar{\nu}_e})$ is taken from ~\cite{IBDcross}. The flux spectrum $\text{d}\phi_{\bar{\nu}_e}^{\rm DM}/\text{d}E_{\bar{\nu}_e}$ is given by~Eq.~(\ref{Neutrino_Flux}). The number of free protons inside the JUNO central detector, $N_{\rm target}$, is about $1.45\cdot 10^{33}$~\cite{JUNO:2015zny}. We set the total exposure time as \SI{10}{years}. The parameter $\epsilon$ is the IBD detection efficiency optimized for the DM search, which is obtained from the official JUNO offline simulation and analysis frameworks. The simulation procedure includes the event generator, detector response simulation, electronics simulation, energy reconstruction, and vertex reconstruction. We apply the same simulation and analysis frameworks to evaluate the backgrounds.
The final efficiency, $\epsilon$, is the product of IBD signal selection, muon veto and PSD cut efficiencies. We note that the cut efficiencies hereafter always represent the event survival probability after the cut, regardless of whether these events are signal or backgrounds.

The criteria for IBD signal selection are composed of the following: \\
(1) cut on the time difference between prompt and delayed signals, $\Delta T$ $<$ \SI{1}{ms}, \\
(2)  \SI{1.8}{\mega\electronvolt} $<$ deposited energy of the delayed signal ($E_{\rm d}$) $<$ \SI{2.6}{\mega\electronvolt}, \\
(3) multiplicity cut condition, $N_{\rm mult} =1$, \\
(4) cut on the root mean square of time residual profile\footnote{To separate atmospheric $\overset{(-)}{\nu_\mu}$ charged-current ($\overset{(-)}{\nu_\mu}$  CC) events from atmospheric $\overset{(-)}{\nu_e}$ charged-current ($\overset{(-)}{\nu_e}$ CC) events and IBD events of other sources (see Sec. 4.3 for further details), we define the time residual $T_{\rm res}$ for each hit on the $i$-th 3-inch PMT as $T^i_{\rm res}=t^i_{\rm hit}-n\cdot R_V^i/c$ with $t^i_{\rm hit}$ the hit time on the $i$-th PMT, $n$ the refraction index of JUNO liquid scintillator (LS) and $R^i_V$ the distance between the reconstructed vertex position and the $i$-th PMT. The time residual profile of the scintillation light emitted by $\overset{(-)}{\nu_\mu}$  CC events is different from that of general $\overset{(-)}{\nu_e}$ CC events including IBD, since $\mu^{\pm}$ from the former takes a longer time to deposit its energy to LS than $e^{\pm}$ from the latter does. Hence the root mean square of the $T_{\rm res}$ distribution over the fired 3-inch PMTs, denoted as $\sigma (T_{\rm res})$, is a useful parameter for event selections~\cite{Atmos_mu_e}. The cut $\sigma(T_{\rm res})$ $<$ \SI{77}{\nano\second} can effectively reject $\overset{(-)}{\nu_\mu}$  CC events.}, $\sigma(T_{\rm res})$ $<$ \SI{77}{\nano\second}, \\
(5) fiducial volume cut, $R<$\SI{16}{\meter},\\
(6) prompt signal deposited energy ($E_{\rm p}$) cut, and \\
(7) prompt-delay distance cut. \\
The efficiencies for (1), (2), (3), and (4) are 99.6\%, 98.2\%, 99.9\%, and 99.9\%, respectively. The efficiency for (5) is 73.9\% owing to the uniform distribution of DM events. 
The efficiency of (6) varies between 92.0\% and 99.8\% for the DM mass range of \SIrange[range-units = brackets, range-phrase=--]{15}{100}{MeV}. The efficiency of (7) is maintained at around 99.5\% by varying the distance cut parameter with $m_{\chi}$. We summarize the efficiencies and the uncertainties in Table \ref{table:DM_cuts}. The muon veto cut efficiency is well studied in JUNO~\cite{Muon_strategies}. In the case that the muon is detected by all detectors except the scintillator detector or when its track is not successfully reconstructed, any event occurring within $\SI{0.2}{seconds}$ after the muon detection is discarded. On the other hand, if the muon is also detected by the central detector and its track is successfully reconstructed, only events within a \SI{3}{\meter} radius cylindrical region around the muon track and occurring within $\SI{0.2}{seconds}$ after the muon are rejected.

\begin{figure}
    \centering
    \includegraphics[width=1\textwidth]{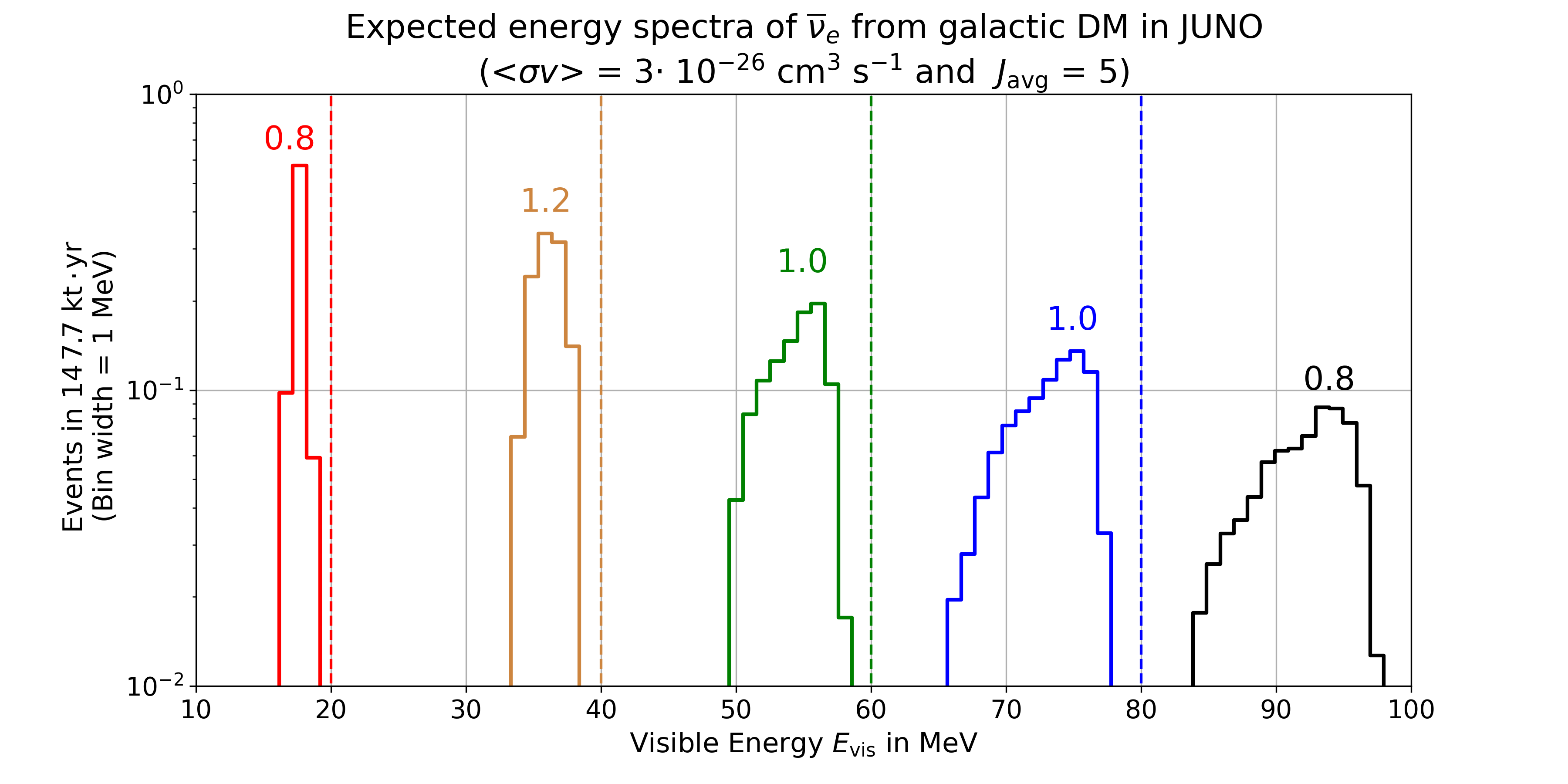}
    \caption{Visible energy spectra in the JUNO detector with \SI{10}{years} of exposure time and \SI{14.77}{\kilo\tonne} of fiducial mass after applying all except the PSD cuts for $J_{\rm avg} = 5$ and $\langle \sigma v \rangle$ = \num{3e-26}\si{\cubic\centi\meter\per\second}. The distributions are shown for
    $m_{\chi}=$ \SI{20}{\mega\electronvolt} (red), \SI{40}{\mega\electronvolt} (brown), \SI{60}{\mega\electronvolt} (green), \SI{80}{\mega\electronvolt} (blue) and \SI{100}{\mega\electronvolt} (black). The total number of signal events is also displayed above the respective peaks.}
    \label{figure_signal_spectrum}
\end{figure}

We present in Fig.~\ref{figure_signal_spectrum} the visible energy spectra of neutrino events induced by DM annihilation. The non-Gaussian appearance of the peaks in Fig.~\ref{figure_signal_spectrum} can be attributed to 
the IBD kinematics and its differential cross section. In our interested energy range, the positron energy is related to its direction by $E_{e^+}\approx (E_\nu-1.30 \ {\rm MeV})\cdot (1-E_\nu(1-\cos\theta)/M)$~\cite{IBDangle} with $\theta$ the positron angle relative to the neutrino direction and $M$ the nucleon mass. The maximal energy of the positron occurs at $\theta=0$ with $E_{e^+}^{\rm max}=(E_\nu-1.30 \ {\rm MeV})$ while the minimal energy corresponds to $\theta=\pi$ so that $E_{e^+}^{\rm min}=(1-2E_\nu/M)\cdot E_{e^+}^{\rm max}$. The relation between $E_{e^+}^{\rm max}$ and $E_{e^+}^{\rm min}$ explains why the DM event spectrum becomes broader as $m_\chi$ ($E_\nu$) increases. Furthermore the positron average energy is given by $\langle E_{e^+}\rangle=(E_{e^+}^{\rm max}+E_{e^+}^{\rm min})/2+E_{e^+}^{\rm max}E_\nu\langle \cos\theta \rangle/M$ with $\langle \cos\theta \rangle\approx 2.4\cdot (E_\nu-13 \ {\rm MeV})/M$~\cite{IBDangle}. Clearly, as $m_\chi$ ($E_\nu$) increases, $\langle E_{e^+}\rangle$ moves farther away from $(E_{e^+}^{\rm max}+E_{e^+}^{\rm min})/2$. This is reflected by the increasingly asymmetrical shape of DM event spectrum as $m_\chi$ increases.  In Sec. 5, we shall discuss the PSD method used for further background suppression. The efficiency for the PSD cut will be evaluated. 
\begin{table}
\centering
\caption[The inverse beta decay selection criteria based on dark matter simulation.]{Summary of IBD signal selection cuts, the muon veto and the PSD cut.}
\label{table:DM_cuts}
\scalebox{0.67}{
\begin{tabular}{llcc}
\hline
\multicolumn{2}{l}{IBD signal selection - $m_{\chi}$ independent }  & cut efficiency & MC uncertainty\\
(1) $\Delta T$ cut,& $\Delta T$ $<$ 1 ms                     & $\sim$99.6$\%$ & 1.0\%\\
(2) $E_{\rm d}$ cut,& 1.8 MeV $<$ $E_{\rm d}$ $<$ 2.6 MeV   & 98.2$\%$ & 1.0\%\\
(3) Multiplicity cut,& $N_{\text{mult}}$ = 1      & 99.9 $\%$ & 1.0\%\\
(4) $T_{\rm res}$ cut,& $\sigma(T_{\rm res}) <$ 77 ns          & 99.9$\%$  & 1.0\%\\
(5) Fiducial volume cut,& $R_{\rm prompt}<$ 16 m    & 73.9 $\%$ & 0.8\%\\
\hline
\multicolumn{2}{l}{IBD signal selection - $m_{\chi}$ dependent }   & cut efficiency,  $15\leq m_{\chi}/{\rm MeV}\leq 100$ & MC uncertainty\\
(6) $E_{\rm p}$ cut &      (0.75 $\cdot$ $m_{\chi}$ + 2) MeV < $E_{\rm p}$ < (0.97 $\cdot$ $m_{\chi}$ - 0.06) MeV       & $\sim$ (99 - 0.04 $\cdot$ $m_{\chi}$/MeV)\%  & 0.8\%\\
(7) $\Delta D$  cut   &      0 mm <  $\Delta D$ < (7.7 $\cdot$ $m_{\chi}$/MeV+226) mm       & $\sim$99.5$\%$ & 1.0\%\\
\hline
Muon veto cut      &                       & $\sim$97.5$\%$ & 1.0\%\\
\hline
PSD cut [15-100] MeV        &               & [30.4$\%$ - 99.9$\%$] & 5.0\%\\
\hline
Total [15-100] MeV        &                         & [22.5$\%$ - 68.8$\%$] & $\sim$4\% \\
\hline \hline
\end{tabular}}
\end{table}

Despite a low event number, the unprecedented energy resolution of JUNO provides a great advantage for observing monochromatic neutrino signatures from DM annihilation as we shall see in Sec. 6. 
%----------- Section 4 Backgrounds ----------
\section{Backgrounds}
In this section, we discuss backgrounds to the indirect DM signature. We divide the backgrounds into two categories: IBD backgrounds and non-IBD backgrounds.

IBD backgrounds come from other $\bar{\nu}_e$ sources, which are intrinsically indistinguishable from $\bar{\nu}_e$ produced by DM annihilation. The reactor neutrinos, charged current interactions of atmospheric neutrinos (Atm-$\nu$ CC) and diffuse supernova neutrino background (DSNB) act as the neutrino floor for the indirect DM search in JUNO. We focus on the energy range beginning at \SI{12}{\mega\electronvolt} since there are a large number of reactor neutrino events for energies less than this, which overwhelm the DM signature by several orders of magnitudes. 

Non-IBD backgrounds mimic the IBD coincidence, including fast neutrons (FN) induced by muons passing through the surrounding rock, radionuclides ($^{11}$Li and $^{14}$B) induced by muon spallation on carbon nuclei, and neutral current interactions of atmospheric neutrinos. These events can be suppressed by proper veto strategies with good efficiencies. Backgrounds from FN are reduced by the fiducial volume cut while the muon spallation backgrounds are reduced by the \SI{12}{\mega\electronvolt} cut on $E_{\rm p}$ and the muon veto cut customized for JUNO. The neutral current interactions from atmospheric neutrinos are suppressed to an acceptable level with the PSD cut. 
%---------------Reactor --------------
\subsection{ Reactor \texorpdfstring{$\bar{\nu}_e$}{neutrino e-bar}}
The JUNO detector is located \SI{52.5}{\kilo\meter} away from the eight reactors of the Taishan and Yangjiang nuclear power plants with a total thermal power of \SI{26.6}{\giga\watt}.
The $\bar{\nu}_e$ flux is generated through $\beta$ decays of $^{235}$U, $^{238}$U, $^{239}$Pu and $^{241}$Pu. We follow ~\cite{JUNO:2015zny} to simulate IBD rate and spectrum from reactor $\bar{\nu}_e$. The total number of IBD events from the reactors is estimated to be $223,736$ after taking into account the oscillation effect for \SI{147.7}{kt\cdot year} exposure. The spectral shape of the reactor neutrino background is derived from Huber-Muller model~\cite{Mueller,Huber}. The event spectrum drops rapidly beyond~\SI{12}{MeV}, resulting in only a few events. To avoid background events from reactor neutrinos, we set the mass range for our DM search to be higher than \SI{15}{MeV}.
%---------------DSNB --------------
\subsection{ DSNB } 
The DSNB is a cumulative neutrino emission generated from core-collapse supernovae with all flavors of neutrinos and anti-neutrinos in the observable universe. The isotropic DSNB flux is given by~\cite{DSNBflux}:
\begin{equation}
{\frac{\text{d}\phi_{\text{DSNB}}}{\text{d}E_{\nu}}} =  \int_0^{z_{\text{max}}}R_{\text{SN}}(z){\frac{\text{d}N_{\nu}(E'_{\nu})}{\text{d}E'_{\nu}}}(1+z)\cdot c\left |\frac{\text{d}t}{\text{d}z}\right|\text{d}z,
\end{equation}
with $z_{\rm max}$ being the maximal redshift to be covered, and $c$ the speed of light. The first term, ${R_{\text{SN}}(z)}$, is the core-collapse supernova (CCSN) rate, which is related to the star formation rate and the initial mass function of the forming stars. The second term, $\text{d}N_{\nu}(E'_{\nu})/\text{d}E'_{\nu}$, is the averaged energy spectrum of the emitted neutrinos per supernova explosion. The energy $E'_{\nu}$ at the source is linked with $E_{\nu}$ observed on the Earth through the redshift relation, $E'_{\nu} = (1+z)E_\nu$. The last term represents the assumed cosmological model, which relates the redshift $z$ to the cosmic time $t$ according to $|\text{d}t/\text{d}z| = 1/(H_0(1+z)\sqrt{\Omega_{m}(1+z)^{3}+\Omega_{\Lambda}})$ with $\Omega_{m}$ being the present-day density parameter of matter, $\Omega_\Lambda$ the fraction of the energy density provided by the dark energy, and $H_0$ the Hubble constant. We take the standard $\Lambda$CDM cosmology with $\Omega_{m}$ = 0.3, $\Omega_{\Lambda}$ = 0.7, and $H_{0}$ = \SI{70}{km\,s^{-1}\,Mpc^{-1}}.

The major uncertainty of the DSNB flux arises from the cosmological SN rate ${R_{\text{SN}}(z)}$ and the average energy spectrum of SN neutrinos $\text{d}N_{\nu}(E'_{\nu})/\text{d}E'_{\nu}$. We adopt the nominal SN rate as $R_{\text{SN}}(0) =$  \SI{1.0e-4}{Mpc^{-3}yr^{-1}}~\cite{RSN} while the range for the SN rate is taken to be \SI{0.5e-4}{Mpc^{-3}yr^{-1}} $\leq R_{\text{SN}}(0) \leq$ \SI{2.0e-4}{Mpc^{-3}yr^{-1}}. The average energy spectrum of SN neutrinos varies with astrophysical parameters such as the explodability of the progenitor, the maximum baryonic neutron star mass, and the cosmic CCSN rate. Therefore, we consider $12\leq \langle E_{\nu} \rangle/{\rm MeV} \leq 18$ for the average energy of SN neutrinos~\cite{E_avg_BHR}. The above-mentioned parameters determine the fraction of failed SN, $f_{\text{BH}}$, which in turn determines $\text{d}N_{\nu}(E'_{\nu})/\text{d}E'_{\nu}$. We adopt $f_{\text{BH}}$ = 27\% as the fiducial model~\cite{E_avg_BHR} and the range for $f_{\text{BH}}$ is taken as $0 \leq f_{\text{BH}} \leq 0.4$ in our study.

The IBD event spectrum induced by the DSNB is given by
\begin{equation}
{\frac{\text{d}N_{\text{DSNB}}(E_{\bar{\nu}_e})}{\text{d}E_{\bar{\nu}_e}}} = \sigma_{\text{IBD}}(E_{\bar{\nu}_e}) \cdot {\frac{\text{d}\phi_{\text{DSNB}}}{\text{d}E_{\bar{\nu}_e}}} \cdot N_{\text{target}} \cdot t \cdot\epsilon_{\text{DSNB}},
\end{equation}
where the total efficiency for the DSNB, $\epsilon_{\text{DSNB}}$, is $47.3\%$ based on IBD signal selection, muon veto and PSD applied for DM search. The low efficiency is largely due to the implementation of PSD, which will be discussed in Section 5. Furthermore, we consider the upper bound of the DSNB flux obtained from the SuperK search~$\cite{DSNB_upperflux_SK}$ as the largest possible DSNB flux. As a result, we obtain the DSNB energy spectrum with an event number ranging from $N_{\rm evts} = 4.2$ given by the low flux model to $N_{\rm evts} = 193.5$ given by the SuperK flux upper bound for \SI{147.7}{kt\cdot year} exposure in the visible energy range of \SIrange[range-units = brackets, range-phrase=--]{12}{100}{MeV}.
\begin{table}
\label{DSNB_evts}
\centering
\caption[The parameters of diffuse supernovae background and event numbers.]{Summary of DSNB flux models used in the background analysis}
\scalebox{0.9}{
\label{table:DSNB_Para}
\begin{tabular}{ccccc}
\hline
flux model & $f_{\rm BH}$ & $\langle E_\nu \rangle$ [MeV]& $R_{\text{SN}}(z) \ [{\rm yr}^{-1} {\rm Mpc}^{-3}]$ &    $N_{\rm evts}$ in (12-100) MeV [147.7 kt $\cdot$ yr]\\
\hline
low    & 0$\%$  & 12 & $0.5\times 10^{-4}$   & 4.2\\
nominal & 27$\%$ & 15 & $1.0\times 10^{-4}$   & $2.3\times 10^1$\\
high   & 40$\%$ & 18 & $2.0\times 10^{-4}$   & $6.6\times 10^1$\\
SuperK &        &    &                       & $1.9\times 10^2$\\
\hline \hline
\end{tabular}}
\end{table}
The relevant parameters for the three flux models and the corresponding DSNB event rates are presented in Table \ref{table:DSNB_Para}.
%---------------CC --------------
\subsection{ Atmospheric \texorpdfstring{$\nu$}{nu} charged current background}
We study the IBD events induced by the atmospheric neutrino charged current interactions in JUNO. 
We adopt the atmospheric neutrino flux calculated by M.~Honda {\it et al.}~\cite{Honda}, which considers all three flavors of neutrinos and their anti-neutrinos, the effect of Earth’s magnetic field on the flux, and neutrino oscillation effects at the JUNO site. The atmospheric neutrino flux below \SI{100}{MeV} is from FLUKA \cite{Battistoni:2015epi} simulation results, which is normalized to match the flux in~\cite{Honda} in the overlapping energy region from \SI{100}{MeV} to \SI{944}{MeV}.
Atmospheric $\nu_{\tau}$ and $\bar{\nu}_\tau$ are neglected in our study due to their low fluxes (both less than 1\% of the total atmospheric neutrino flux), while the remaining $\nu_{\mu}$ ($\bar{\nu}_{\mu}$) and $\nu_e$ ($\bar{\nu}_e$) fluxes are comparable. These two flavors of  atmospheric neutrino fluxes can be separated with an accuracy better than 99.9\% in the JUNO detector for the energy range of \SIrange[range-units = brackets, range-phrase=--]{12}{100}{MeV}~\cite{Atmos_mu_e}. The background due to atmospheric $\nu_{\mu}$ and $\bar{\nu}_{\mu}$ can be suppressed through a $T_{\rm res}$ cut that is applied on the root mean square of the time residual profile as mentioned in Sec. 3. The signal selection efficiency with the $T_{\rm res}$ cut remains at $99.98\%$. Therefore we focus on backgrounds due to atmospheric $\nu_e$ and $\bar{\nu}_e$. 
We have simulated $\bar{\nu}_e+p$ and $\nu_e  (\bar{\nu}_e) +^{12}\text{C}$ interactions with GENIE (2.12.0)~\cite{GENIE} and the JUNO offline framework, which produces results consistent with previous studies on neutrino-nucleus interactions~\cite{12C_Xsec}. The IBD channel, $\bar{\nu}_e + p \rightarrow e^+ + n$, gives the dominant event rate, 30.5 $\pm$ 7.6 for \SI{147.7}{kt\cdot year} for the visible energy range of \SIrange[range-units = brackets, range-phrase=--]{12}{100}{MeV}. The second background channel, $\bar{\nu}_e+^{12}\text{C} \rightarrow ^{12}\text{B}^*+e^+$ with the secondary decay $^{12}\text{B}^*\rightarrow^{11}\text{B}+n$, gives the next-to-leading event rate for the same exposure and energy range. This channel can mimic IBD events with a prompt $e^+$ and a delayed neutron coming from the secondary $^{12}\text{B}^*$ decay. 
Due to comparable event rates with the signal, the two channels, $\bar{\nu}_e + p \rightarrow e^+ + n$ and $\bar{\nu}_e + ^{12}\text{C} \rightarrow e^+ + n + ^{11}\text{B}$, will be included in the following pulse shape and sensitivity analyses.  
%---------------Muon spallation --------------
\subsection{ Cosmogenic isotopes}
The $\beta$-n decay from isotopes could mimic IBD events by emitting a $\beta$-particle ($e^-$ or $e^+$) and a neutron. This effect has been measured by both KamLAND~\cite{Isotope_KamLAND} and Borexino~\cite{Isotope_Borexino}. The isotopes could be produced in the JUNO site through interactions between energetic cosmic muons and $^{12}$C. 
We only focus on isotopes giving rise to $\beta$-n decays with significant event rates and $Q$ values higher than \SI{10}{MeV}. Using FLUKA, the rates of  $^9\text{Li}$, $^{11}\text{Li}$, $^{12}\text{Be}$ and $^{14}\text{B}$ are presented in Table~\ref{table:MuIso}~\cite{JUNO:2015zny}. The background from $^9\text{Li}$ and $^{12}\text{Be}$ can be neglected due to a \SI{12}{MeV} cut on the prompt energy, which is the lower energy limit of our DM search. The half-lives of $^{11}\text{Li}$ and $^{14}\text{B}$ are \SI{8.75}{ms} and \SI{12.6}{ms}, respectively. They both contribute to $\beta$-n decays with branching fractions of $83\%$ and $6.1\%$, respectively. For the total event rate of $\rm {^{11}Li}$ and $\rm {^{14}B}$, we assume a 10\% uncertainty based on the Poisson error. A flat spectrum is assumed for the event rate estimation. As a result, the total event rate of the $\beta$-n decays is 57.0 $\pm$ 5.7 for \SI{147.7}{kt\cdot year} for the visible energy above \SI{12}{MeV}.

Muon veto strategies in JUNO have been studied in~\cite{Muon_strategies}. To suppress $^{11}\text{Li}$ and $^{14}\text{B}$ background events, we choose \SI{0.2}{s} as the deadtime interval, i.e., after each muon event a \SI{0.2}{s} of exposure time is ignored. Due to this veto, the exposure efficiency in JUNO becomes $97.5\%$ while more than $99\%$ of $^{11}\text{Li}$ and $^{14}\text{B}$ events are vetoed. Therefore, after applying the muon veto and the \SI{12}{MeV} energy cut, $^9\text{Li}$, $^{11}\text{Li}$, $^{12}\text{Be}$ and $^{14}\text{B}$ events are negligible.
%---------------Muon induced fast neutron --------------
\subsection{ Muon induced fast neutrons}
\begin{figure}
\centering
\includegraphics[scale=0.25]{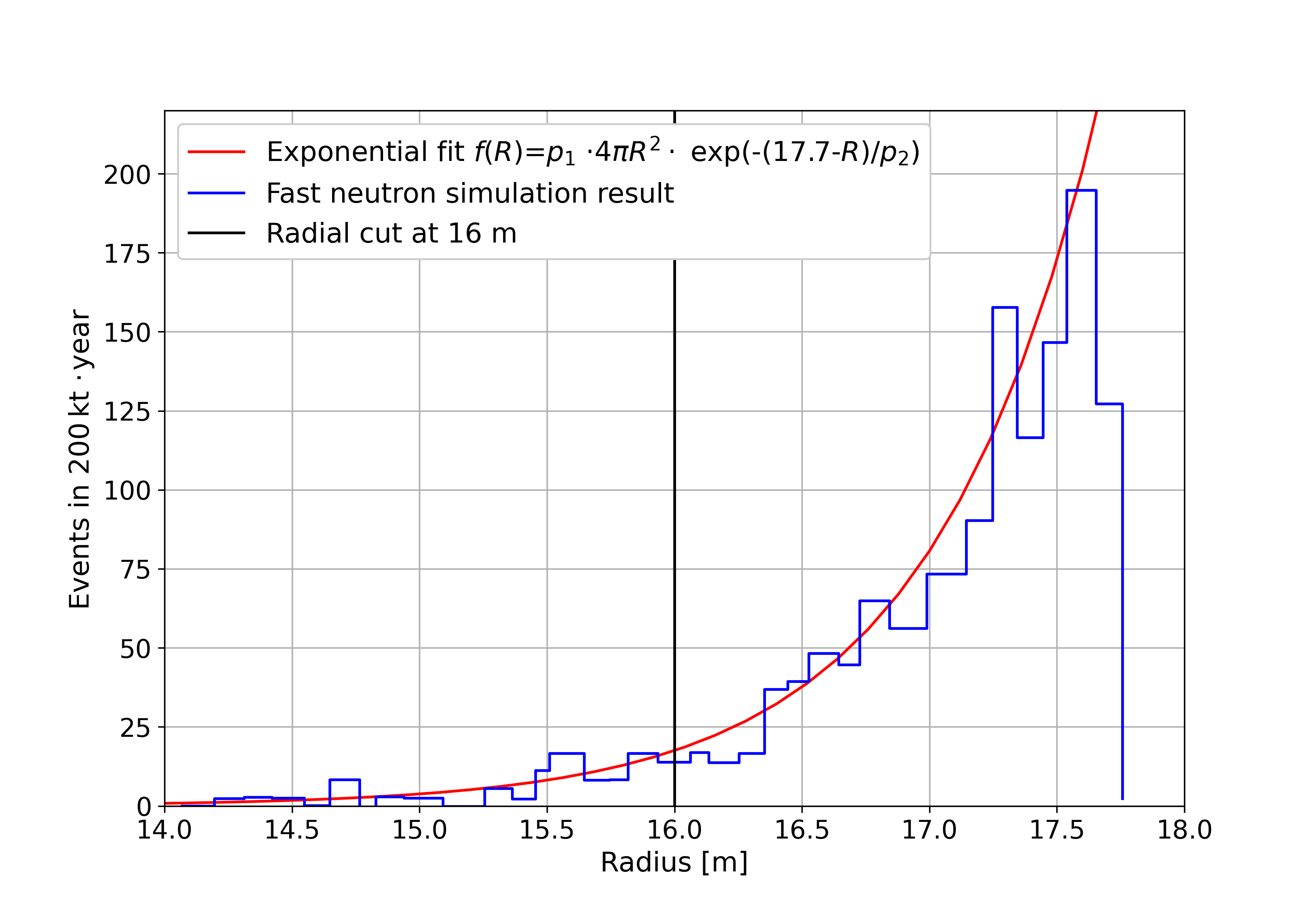}
\caption[Radius distribution of muon-induced FN events.]{Histogram: radial distribution of muon-induced FN events. Red line: fit to the histogram with the function $f(R)=p_1\cdot4\pi R^2\cdot\exp (-(17.7 - R)/p_2)$ with $p_1$=$0.75\pm 0.04$,
and $p_2$$= 0.71\pm 0.04$. Black line: radial cut at \SI{16}{m}.}
\label{fastN}
\end{figure}

The above-mentioned muon veto tags muons passing through the LS and the water buffer with excellent efficiencies of 100\% and 99.8\%, respectively~\cite{Muon_strategies}. However, untagged muons that are either corner clipping the detector or passing through the rocks surrounding the detector could produce energetic neutrons. Any of these neutrons may enter in the LS and produce the prompt proton-recoil signal before being captured by the hydrogen, which is called a fast neutron (FN) event. These signatures mimic IBD events.
\begin{table}
\label{Mu_Iso}
\centering
\caption[The muon isotopes event that mimics IBD]{The estimated rates of cosmogenic isotopes in JUNO for energies above 10 MeV. \\}
\scalebox{0.85}{
\label{table:MuIso}
\begin{tabular}{c|c|c|c|c}
isotope & decay mode & $Q$ in MeV & half-life $T_{1/2}$ &  Rate per 10 years ($E>10$ MeV)
\\
\hline
$^{9}{\rm Li}$  & $e^- + n$ & 11.9  & 178 ms   & $2.8\times 10^4$\\
$^{11}{\rm Li}$ & $e^- + n$ & 20.6  & 8.75 ms  & $9.3\times 10^1$\\
$^{12}{\rm Be}$ & $e^- + n$ & 11.7 & 21.5 ms  & 1.2\\
$^{14}{\rm B}$  & $e^- + n$ & 20.6 & 12.6 ms  & 2.4\\
\end{tabular}}
\end{table}
The FN simulation has been performed based on the JUNO official simulation framework. The simulated atmospheric muon event sample corresponds to around \SI{1200}{days} of data taking. A flat FN energy spectrum is obtained and the event rate is  
1340 $\pm$ 270 for \SI{200}{kt\cdot year} 
of exposure within the visible energy range of \SIrange[range-units = brackets, range-phrase=--]{12}{100}{MeV} of the prompt signal. The spatial vertex distributions of FN events are shown in Fig.~\ref{fastN}. 
The FN event rate decreases from the edge to the center of the detector, which results from the neutron attenuation.    

We perform the fiducial volume cut at the radius $R=$\SI{16}{m}, which significantly reduces the FN event rate. Using an exponential fit to the simulated FN data, we estimate the fiducial volume cut efficiency to be 7.1\%. In contrast, the fiducial volume cut efficiency for the uniformly distributed events, such as DM, DSNB, and atmospheric neutrino events is $73.9\%$. The FN events with $R<$\SI{16}{m} can be suppressed by the PSD method, which will be discussed in the next section. Eventually, these FN backgrounds are negligible.
%---------Atmospheric neutral current events------------
\subsection{ Atmospheric \texorpdfstring{$\nu$}{nu} neutral current background}
A detailed study on the neutral-current interaction between the atmospheric neutrino and $^{12}\text{C}$ has been carried out in~\cite{NC_12C_AtmNu-I}. We have reproduced the above simulations and applied them to the DM study in the JUNO offline framework. 
We have neglected the channel $\nu+p\to \nu+p$ since the fraction of neutron-less IBD-like events from atmospheric neutrino NC interactions is only $0.99\%$. 

Using the atmospheric neutrino flux calculated by M.~Honda {\it et al.}~\cite{Honda}, we estimate the Atm-$\nu$ NC background. The uncertainty of this flux calculation is less than 10\% in the energy range of \SIrange[range-units = brackets, range-phrase=--]{1}{10}{GeV} while it varies between (10-30)\% outside this energy range due to the lack of observational results. We use the neutrino generator GENIE (2.12.0)~\cite{GENIE} to model neutral current interactions between atmospheric neutrinos and $^{12}\text{C}$. We adopt the default setting in GENIE, where the axial mass $M_A$ in the parametrization of the nucleon axial-vector form factor is taken as \SI{0.99}{GeV}, the relativistic Fermi gas (RFG) model is adopted for nuclear structures, and the Intranuclear Cascade (INC) model is applied for final-state interactions~\cite{GENIE_FSI}. The de-excitation of the final-state nuclei is simulated with the package TALYS (1.8)~\cite{TALYS}. A statistical configuration model~\cite{12C_deexc_model1,12C_deexc_model2,12C_deexc_model3} is applied for providing the de-excitation probability before the TALYS simulation.

We conclude that the most important NC interaction channel is $\nu+^{12}\text{C}\to \nu+^{11}\text{C}+n$, and the total event rate before IBD signal selection is estimated to be \SI{49.0}{year^{-1} kt^{-1}}, which is consistent with the previous study~\cite{NC_12C_AtmNu-I}. Applying IBD event selection criteria, the efficiency for Atm-$\nu$ NC event becomes (7.3 $\pm$ 0.5)\%. A conservative 15\% total uncertainty is taken from the study in~\cite{NC_12C_AtmNu-II}. Hence the event rate of IBD-like atmospheric neutrino NC events is 
670 $\pm$ 100 for \SI{147.7}{kt\cdot year} in the visible energy range \SIrange[range-units = brackets, range-phrase=--]{12}{100}{MeV}.
%\textcolor{red}{665 $\pm$ 103} for \SI{147.7}{kt\cdot year} in the visible energy range \textcolor{red}{\SIrange[range-units = brackets, range-phrase=--]{12}{100}{MeV}}. 
%-----------Section 5 PSD----------
%---------Pulse Shape Analysis------------
\section{ Pulse Shape Discrimination }
Different types of particles show distinct photon emission time profiles that result from LS excitation induced by the deposited energy. Pulse shape discrimination (PSD) is a powerful way to separate Atm-$\nu$ NC and FN events from IBD signal events by analyzing the pulse shapes of their prompt signals. 
The tail-to-total ratio (TTR) method~\cite{LENA_T2T} is adopted here, where the ratio between the charge in a specific time window corresponding to the tail of the pulse and the total charge of the pulse is the parameter to distinguish between different event types. Although multivariate analysis (TMVA) \cite{TMVA} and machine learning methods~\cite{Skitlearn} show superior efficiencies, the TTR method is sufficient for suppressing Atm-$\nu$ NC events to the same order as the neutrino floor in our DM study. 

\begin{figure}
\centering
\includegraphics[scale=0.17]{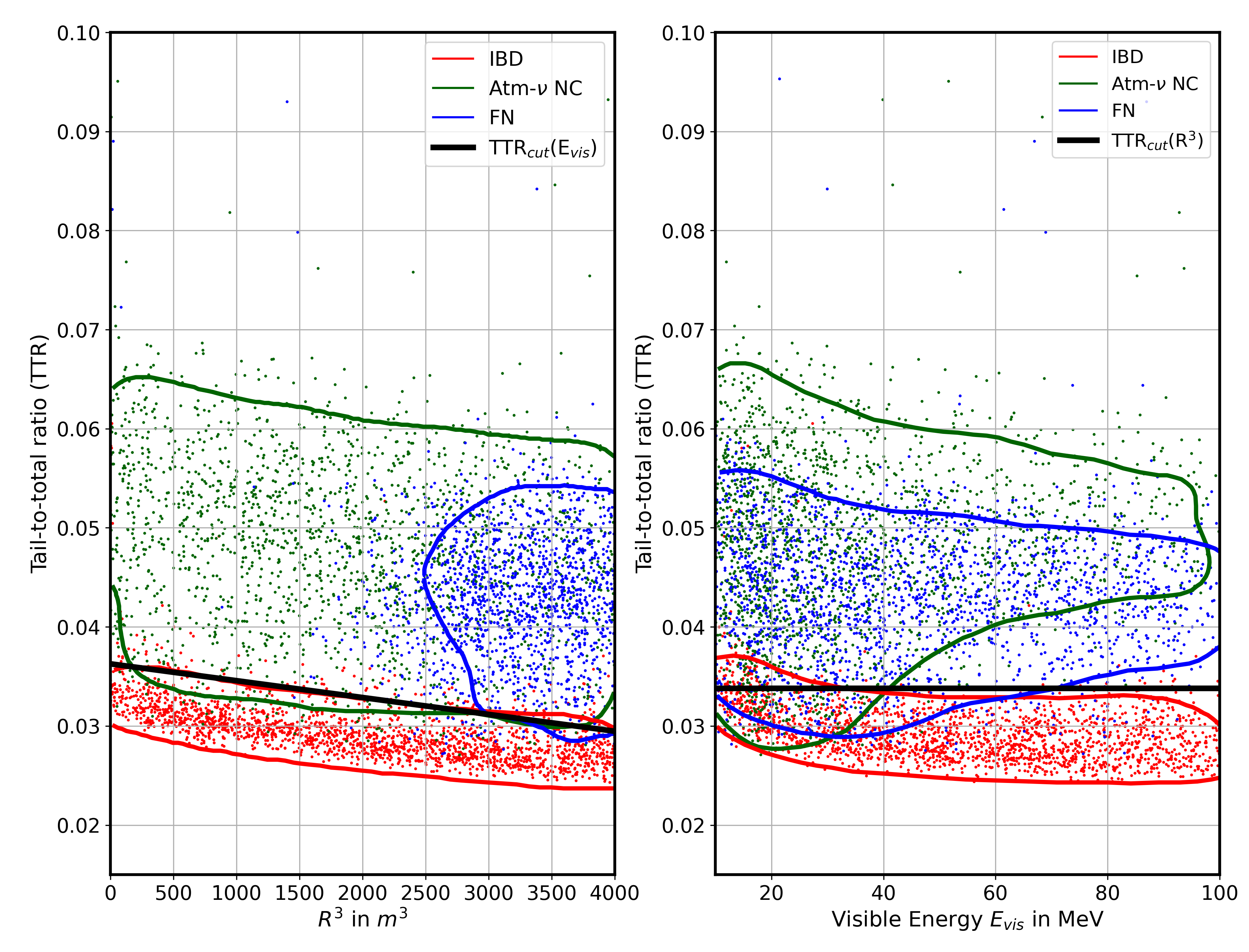}
\caption[Tail-to-total ratios with PSD selection criteria.]{Tail-to-total ratios (TTR) versus visible energy and the reconstructed position for IBD events, Atm-$\nu$ NC and FN events that pass
the IBD selection criteria. Different event types are represented by scatter plots with different colors. Each contour marks the region that contains 90\% of a given type of events. The black solid curve on each panel is the signal selection criterion such that those events situated below this curve are classified as signals. On the left panel the curve is parametrized by $\rm{TTR_{cut}}$($R$) = 0.036 - 1.7·10$^{-6}$ m$^{-3}\cdot$ $R^3$ while the curve on the right panel is parametrized by $\rm{TTR_{cut}}$($E_{\rm vis}$) = 0.034. The PSD cut selects only those events which are situated below both curves.}
\label{T2T}
\end{figure}

\begin{figure}
\centering
\includegraphics[scale=0.27]{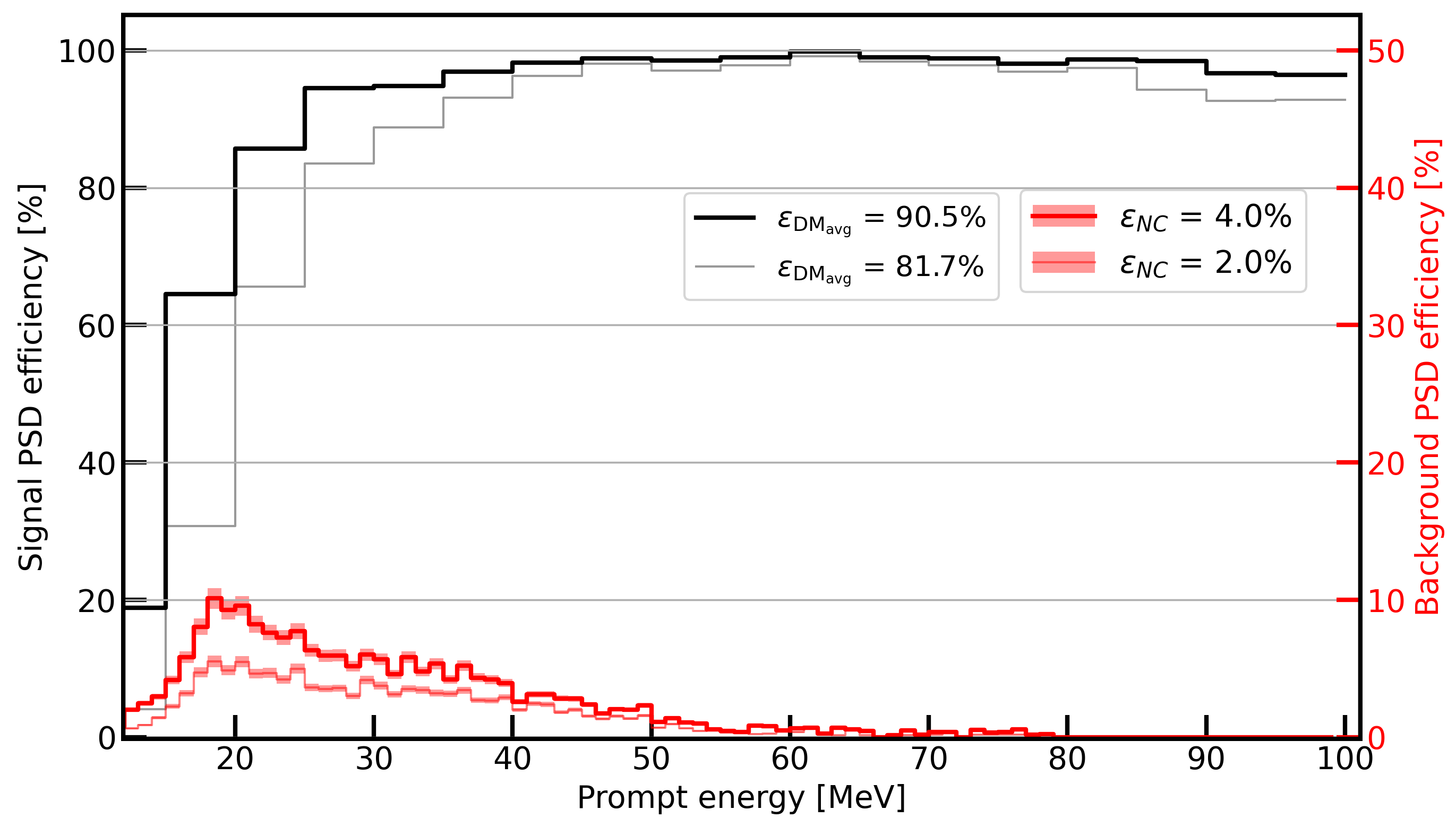}
\caption[The veto efficiencies of pulse shape discrimination apply on DM, Atmos NC and Muon FN events.]{PSD cut efficiencies as functions of the visible energy of the prompt signal of Atm-$\nu$ NC background events (in red) and IBD signal events (in black). Two examples, $\epsilon_{\rm PSD,NC} = 2~\%$ with $\epsilon_{\rm PSD,IBD} = 81.7~\%$ (light curves), and $\epsilon_{\rm PSD,NC} = 4~\%$ with $\epsilon_{\rm PSD,IBD} = 90.5~\%$ (bold curves), are presented.}
\label{PSDeff2d}
\end{figure}

We perform a full simulation to produce pulse shapes of different event types, which is based on the official JUNO offline simulation and analysis frameworks.
To do this, we have employed the DM flux model discussed in Sec. 3, the DSNB flux model discussed in Sec. 4.2, the Atm-$\nu$ CC and NC events from the GENIE interaction models and the FN background events. In our analysis, we select all events that pass the IBD signal selection and muon veto cut. We scan through the tail settings with \SI{50}{ns} steps on the start and end times of the tail. The optimized tail window is from \SI{200}{ns} to \SI{600}{ns} (see discussions later). The TTR ratio versus the visible energy and the reconstructed position are presented via scatter plots in Fig.~\ref{T2T}. There are also contours corresponding to different event types. Each contour marks the region that contains $90\%$ of a given type of events. DM signal and DSNB events occurring through the IBD process result into an identical TTR ratio distribution. This is a relatively stable distribution as reflected by its smooth $90\%$ event boundary. FN events follow from neutron elastic scattering with proton or $^{12}$C and producing gammas, which give rise to a more complex TTR ratio distribution. Finally, Atm-$\nu$ NC events result from neutral-current interactions between atmospheric neutrinos and $^{12}$C, which produces the most complicated TTR ratio distribution due to the variety of  interaction channels characterized by different prompt-signal spectral shapes. 
The black solid curve on each panel represents the event selection criterion such that those events situated below this curve are classified as signals. It is seen that the IBD contour region is entirely below this curve on the left panel, while a small part of the IBD contour region is above the event selection curve for $E_{\rm vis}\leq$ \SI{30}{MeV}. This implies that the DM signal efficiency after the PSD cut remains higher than $90\%$ for $E_{\rm vis}\geq$ \SI{30}{MeV}. On the other hand, for $E_{\rm vis}<$ \SI{30}{MeV}, a significant fraction of IBD events is also removed by the PSD cut. As will be discussed later, such an event reduction is seen for DM signals illustrated by Fig.~\ref{PSDeff2d} and DSNB events illustrated by Figs.~\ref{spectrum_a} and \ref{spectrum_b}.     

We evaluate the PSD performance based on the signal to background ratio $N_{S}/\sqrt{N_{S}+N_{B}}$ averaged over the DM mass range of \SIrange[range-units = brackets, range-phrase=--]{15}{100}{MeV} (with a \SI{5}{MeV} step size), where $N_{S}$ is the number of signal events from DM annihilation for a specific DM mass and $N_{B}$ is the sum of all backgrounds. 
The current best limit on the thermally averaged DM annihilation rate set by SuperK~\cite{SuperK_782} is adopted here to represent the highest allowed signal to background ratio under the latest constraints. 
The PSD efficiencies of IBD events giving rise to the best signal to background ratio are analyzed while keeping the PSD efficiency of NC background events fixed at 2\%, 3\%, ..., 6\%, and 7\%, respectively. In Table \ref{table:PSD eff}, we present the tail window and the resulting PSD efficiency of IBD events which optimize the signal-to-background ratio for a fixed NC-background PSD efficiency. Fig.~\ref{PSDeff2d} illustrates PSD efficiencies as functions of the prompt energy in two different cases, $\epsilon_{\rm PSD,NC} = 2\%$ and $4\%$, respectively. We adopt the setting with 4\% PSD efficiency for NC background events since it gives rise to a better signal-to-background ratio.
\begin{table}
\label{PSD_eff}
\centering
\caption[The pulse shape discrimination efficiencies.]{PSD cut efficiencies for Atm-$\nu$ NC, IBD and FN events, corresponding tail windows and the resulting best mean signal-to-background ratios.}
\label{table:PSD eff}
\begin{tabular}{cccccc}
\hline
$\epsilon_{\rm PSD,NC}$ & $\epsilon_{\rm PSD,IBD}$ &  $\epsilon_{\rm PSD,FN}$ & Tail window (ns) & mean $N_{S}/\sqrt{N_{S}+N_{B}}$ \\
\hline
2.0\% & 81.7\%  &  0.5\%&  [200,600]     & 4.092       \\
3.0\% & 87.5\%  &  1.2\%&  [250,600]   & 4.140       \\
\textbf{4.0\%} & \textbf{90.5\%}  &  \textbf{2.0\%} &  \textbf{[200,600]}  & \textbf{4.151}      \\
5.0\% & 92.3\% &  3.9\% &  [200,600]  & 4.120      \\
6.0\% & 93.6\%  &  7.0\%&  [150,600]  & 4.071      \\
7.0\% & 95.5\%  &  8.2\%&  [200,600]  & 4.054      \\

\hline \hline
\end{tabular}
\end{table} 

As shown in Figs.~\ref{T2T} and ~\ref{PSDeff2d}, PSD is a powerful tool to discriminate IBD events against Atm-$\nu$ NC events, especially for visible energies above \SI{30}{MeV}. Below \SI{30}{MeV}, the discrimination power decreases because of the lower photon statistics. Furthermore, there is a large fraction of $\gamma$'s emitted in the de-excitation processes of Atm-$\nu$ NC events in this energy range. This also reduces the PSD efficiency by realizing that $\gamma$'s from prompt signals of Atm-$\nu$ NC events and positrons from prompt signals of IBD signal events give rise to almost identical photon emission time profiles, which makes the PSD method ineffective. 
Finally, we note that the PSD efficiency might further worsen at higher energies due to energy dependencies of pulse shapes, which require careful studies.

To determine the uncertainty of the PSD cut, we note that spallation neutrons have been proposed as ideal sources for such a study since they induce prompt and delayed pairs with energies similar to those of Atm-$\nu$ NC events. Specifically, around \SI{180}{days} of muon simulation data was used for generating the aforementioned neutron sample~\cite{JUNO_DSNB}, so that the number of spallation neutrons in the signal energy window can be estimated for \SI{10}{years} of data taking. 
Taking an average PSD efficiency for Atm-$\nu$ NC events as $4\%$, which applies to the spallation neutron sample as well, the statistical uncertainty for the selected spallation neutrons are determined by the number of selected neutrons, which varies with the signal energy window determined by $m_{\chi}$ (see Fig.~\ref{figure_signal_spectrum}). It is $16\%$, $10\%$, $15\%$, $24\%$, and $43\%$ with \SI{10}{years} of data taking for $m_\chi=15$, $20$, $30$, $40$, and \SI{50}{MeV}, respectively. 
We stress that the large uncertainty in the high energy range does not cause a huge impact on the DM sensitivity because Atm-$\nu$ NC events do not dominate the total spectrum over \SI{40}{MeV} after applying the PSD cut. Last but not least, several calibration sources and techniques, such as AmBe and Michel electrons, were also proposed for further constraining the PSD systematic in the future~$\cite{JUNO_Calibration}$. \\

In conclusion, we obtain an average PSD efficiency of 90.5\% for IBD signal events from DM annihilation and an average PSD survival probability of 4.0\% for Atm-$\nu$ NC events, as well as approximately 50\%, 18\%, 97\% and 2.0\% for DSNB events from model predictions, DSNB events corresponding to the SuperK upper bound, Atm-$\nu$ CC, and FN events, respectively.
%-----------Section 6 Sensitivity----------
\section{Sensitivity}
The JUNO sensitivity to the detection of $\bar{\nu}_e$ from DM annihilation in the galactic halo will be discussed in this section. First, we summarize the DM signal and backgrounds with the corresponding veto methods applied. Two different approaches, Poisson-type log-likelihood ratio test and Bayesian analysis, are employed to calculate the DM detection sensitivities and verify the consistency of the two approaches.
\subsection{Total spectrum in JUNO}
\begin{figure}
\centering
\includegraphics[width=0.85\textwidth]{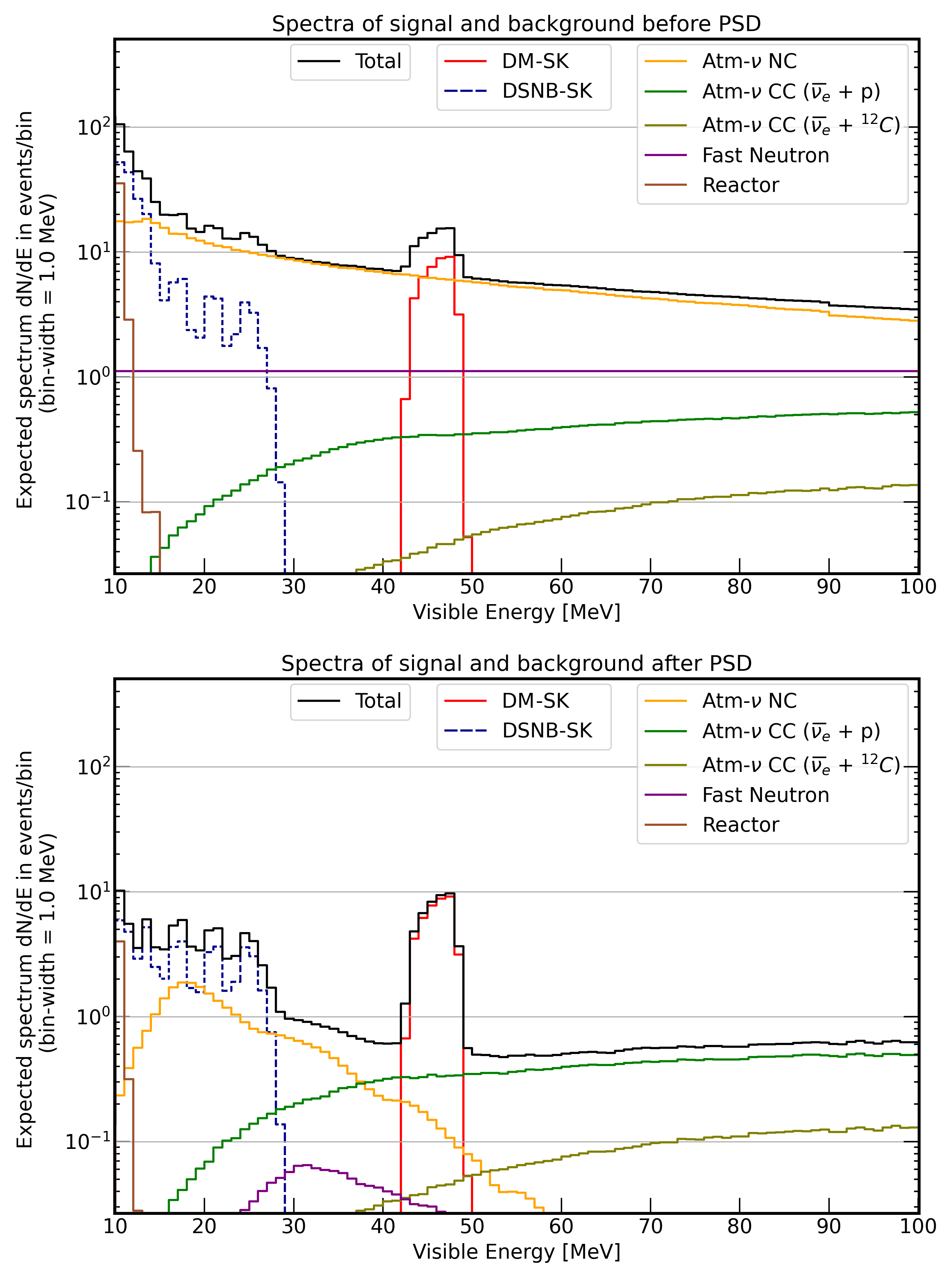}
\caption[Total signal and background spectra in JUNO (with and without PSD methods).]{Total visible energy spectra of DM-IBD signals ($m_{\chi}=$\SI{50}{MeV}) and backgrounds in JUNO before (upper panel) and after PSD (lower panel). Here DSNB~\cite{DSNB_upperflux_SK} and DM~\cite{SuperK_782} event spectra are taken from experimental upper bounds by SuperK.}
\label{spectrum_a}
\end{figure}

\begin{figure}
\centering
\includegraphics[width=0.85\textwidth]{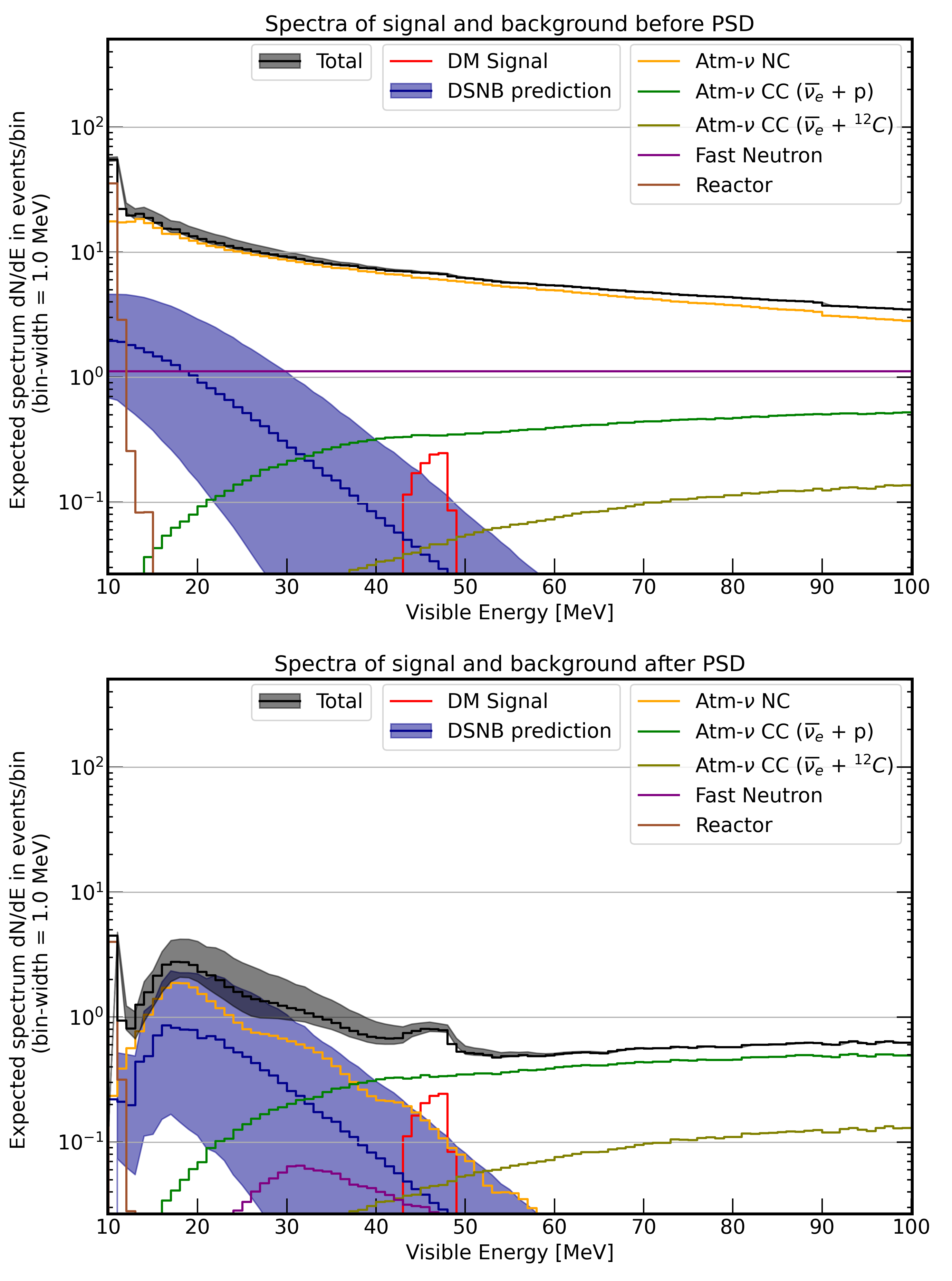}
\caption[Total signal and background spectra in JUNO (with and without PSD methods).]{Same as Fig.~\ref{spectrum_a} with DSNB  event spectra given by theoretical predictions and the DM signal event spectrum based upon thermal relic DM annihilation rate.}
\label{spectrum_b}
\end{figure}

We adopt \SI{10}{years} of exposure time as a reasonable time scale. Canonical value, $J_{\rm avg}= 5$ \cite{Palomares-Ruiz:2007trf}, is used in our analysis. The thermal relic DM annihilation rate, $\langle \sigma v \rangle=$\SI{3e-26}{\cubic\centi\meter\per\second}, is assumed, which results in few signal events for an exposure of \SI{147.7}{kt\cdot year}. 
Figs.~\ref{spectrum_a} and \ref{spectrum_b} show the final visible energy spectra of the signal and backgrounds before (upper panel) and after (lower panel) the PSD cut. DSNB and DM fluxes in Fig.~\ref{spectrum_a} are taken from SuperK flux upper bounds~\cite{DSNB_upperflux_SK, SuperK_782} while those in Fig.~\ref{spectrum_b} are given by theoretical predictions. The signal spectrum is shown for an assumed DM mass of \SI{50}{MeV}.
It is obtained with IBD signal selection cuts, the muon veto cut, and the PSD cut applied (lower panels). With the PSD cut, non-IBD backgrounds (Atm-$\nu$ NC and FN) can be suppressed to the similar order of magnitude as IBD backgrounds (atmospheric CC, reactor, and DSNB). The IBD backgrounds and Atm-$\nu$ NC events build up the background floor and dominate the JUNO sensitivity of indirect DM search. In Table \ref{table::evt}  we summarized the event numbers with IBD signal selection, muon  veto, and PSD cut.  
Despite the low event rate, the mono-energetic $\bar{\nu}_e$ flux yields a sharp visible energy spectrum for the signal, which can be identified against the background spectra. 
\begin{landscape}
\begin{table}
\centering
\fontsize{7.5}{9}\selectfont
\setlength{\tabcolsep}{2pt}

\begin{longtable}{c|c|c|c|c||c|c|c|c||c|c|c|c|c|c|c}
\caption{Summary of DM signal and background event numbers with \SI{147.7}{kt\cdot year} of exposure in JUNO.}
\label{table::evt}\\
\hline
\hline
\textbf{DM}, $\langle\sigma v\rangle$=\SI{3e-26}{\cubic\centi\meter\per\second}, $J_{\rm avg}$ = 5  & \multicolumn{4}{c||}{$\textbf{Event number, IBD signal selection}$}  & \multicolumn{4}{c||}{$\textbf{muon veto}$~\cite{Muon_strategies}}  & \multicolumn{4}{c||}{$\textbf{muon veto}$~\cite{Muon_strategies}+$\textbf{PSD cut}$} & \multicolumn{1}{c}{$\textbf{Signal to Background Ratio}^1$}\\ 
\hline
15 MeV     & \multicolumn{4}{c||}{1.9}  & \multicolumn{4}{c||}{1.8}  & \multicolumn{4}{c||}{0.6} & \multicolumn{1}{c}{0.32}\\
20 MeV     &\multicolumn{4}{c||}{1.5}  & \multicolumn{4}{c||}{1.4}  & \multicolumn{4}{c||}{0.8} & \multicolumn{1}{c}{0.37} \\
90 MeV    & \multicolumn{4}{c||}{0.9}  & \multicolumn{4}{c||}{0.9}  & \multicolumn{4}{c||}{0.9} & \multicolumn{1}{c}{0.27}\\
100 MeV   &  \multicolumn{4}{c||}{0.8}  & \multicolumn{4}{c||}{0.8}  & \multicolumn{4}{c||}{0.8} & \multicolumn{1}{c}{0.26}\\
\hline
$\textbf{Backgrounds}$ & \multicolumn{4}{c||}{}  & \multicolumn{4}{c||}{}  & \multicolumn{4}{c}{} \\
\hline
Cosmogenic Isotopes & \multicolumn{4}{c||}{($0.6 \pm 0.1) \cdot 10^2$} & \multicolumn{4}{c||}{negligible} & \multicolumn{4}{c}{negligible} & \multicolumn{1}{c}{} \\
Fast neutrons  &  \multicolumn{4}{c||}{($1.0 \pm 0.2) \cdot 10^2$} & \multicolumn{4}{c||}{($0.9 \pm 0.2) \cdot 10^2$}  & \multicolumn{4}{c}{($0.2 \pm 0.1) \cdot 10^1$} & \multicolumn{1}{c}{}\\
Atmos NC      & \multicolumn{4}{c||}{($6.7 \pm 1.0) \cdot 10^2$} & \multicolumn{4}{c||}{($6.5 \pm 1.0) \cdot 10^2$} &  \multicolumn{4}{c}{($2.6 \pm 0.7) \cdot 10^1$} & \multicolumn{1}{c}{}\\
Atmos CC      &  \multicolumn{4}{c||}{($0.3 \pm 0.1) \cdot 10^2$} & \multicolumn{4}{c||}{($0.3 \pm 0.1) \cdot 10^2$}  & \multicolumn{4}{c}{($2.9 \pm 0.8) \cdot 10^1$} & \multicolumn{1}{c}{}\\
%\hline
DSNB [ SK $\vert$ Max $\vert$ Nom. $\vert$ Min ]&  $1.9\cdot 10^2$ & $0.7\cdot 10^2$ & $0.2\cdot 10^2$ & 4.2 & $1.9\cdot 10^2$ & $0.6\cdot 10^2$ & $0.2\cdot 10^2$ & 4.1  & $3.4\cdot 10^1$ & $3.5\cdot 10^1$ & $1.1\cdot 10^1$ & \multicolumn{1}{c}{$0.1\cdot 10^1$} & \multicolumn{1}{c}{} \\
\cline{1-13}
$\textbf{Total background}$ & $10.9\cdot 10^2$ & $9.7\cdot 10^2$ & $9.2\cdot 10^2$ & $9.0\cdot 10^2$ & $10.1\cdot 10^2$ & $8.8\cdot 10^2$ & $8.4\cdot 10^2$ & $8.2\cdot 10^2$ & $9.3 \cdot 10^1$ & $9.4 \cdot 10^1$ & $7.0 \cdot 10^1$ & \multicolumn{1}{c}{$6.0 \cdot 10^1$} & \multicolumn{1}{c}{} \\
\cline{1-13}
Total background uncertainty     &  \multicolumn{4}{c||}{$\pm 1.0 \cdot 10^2$} & \multicolumn{4}{c||}{$\pm 1.0 \cdot 10^2$}  & \multicolumn{4}{c}{$\pm 1.1 \cdot 10^1$}  \\
\hline 
\hline 
\end{longtable}
\begin{tablenotes}
            \item[1] $^1$The nominal DSNB flux is adopted in the computation of  signal-to-background ratio.
        \end{tablenotes}
\end{table}

\end{landscape}
\subsection{Two approaches to the sensitivity}
\begin{figure}
\centering
\includegraphics[width=1.0\textwidth]{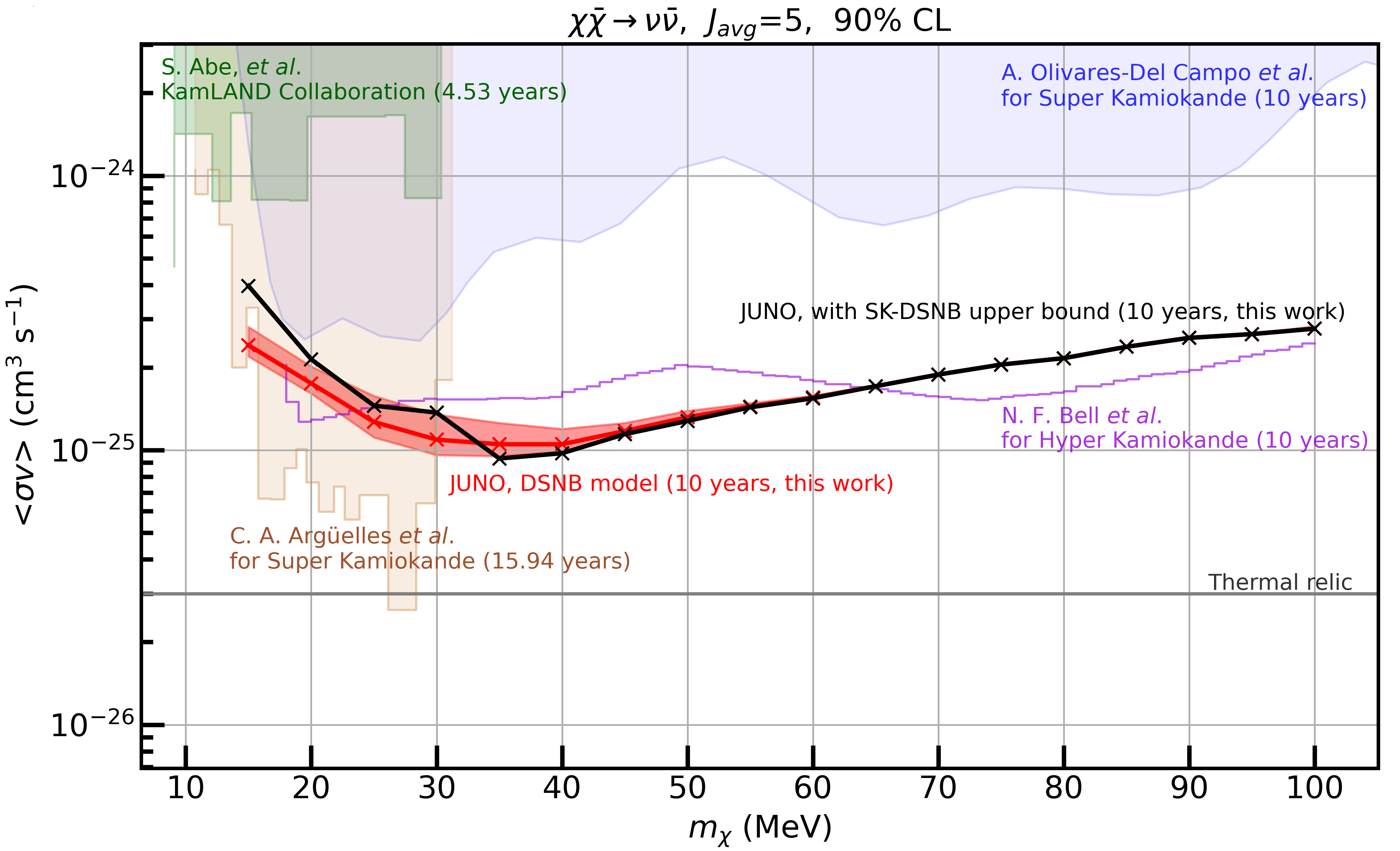}
\caption[90\% confidence level of thermally average cross section in JUNO and other detectors.]{90\% C.L. upper limits on $\langle \sigma v\rangle$ with \SI{10}{years} of data taking in JUNO. Black and red lines represent upper limits with the theoretical model and experimental limit of DSNB, respectively. Due to consistent results between the Bayesian analysis and Likelihood-ratio test, one single upper limit curve for each DSNB flux setting is presented here. Colored areas are the excluded parameter regions from KamLAND~\cite{KamLAND} and SuperK~\cite{SuperK_782,SuperK_1542} observations. HyperK expectation by N.~F.~Bell {\it et al.}~\cite{HyperK} is also shown for comparison. The thermal relic value  $\langle \sigma v\rangle=$\SI{3e-26}{\cubic\centi\meter\per\second} is also shown.}
\label{Upperlimit}
\end{figure}

In the following we apply a Likelihood-ratio test and a Bayesian analysis to estimate the $90\%$ confidence level upper limit on $\langle \sigma v \rangle$ expected by JUNO for the DM mass range of \SIrange[range-units = brackets, range-phrase=--]{15}{100}{MeV}.
\paragraph{Likelihood-ratio test method }
In the likelihood-ratio test method, we define
\begin{equation}
\chi^2= -2\ln \lambda = 2\sum_{i=1}^{N}{({n_i}\ln\frac{n_i}{v_i}+v_i-n_i)} + \sum_{i=1}^{N}\left(\frac{v_i-\bar{v}_i}{\sigma_i}\right)^2,
\end{equation}
where $\lambda$ is the likelihood ratio, $n_i$ represents observed (signal plus background) events per bin, $v_i$ represents background events per bin with its central value denoted by $\bar{v}_i$, and $\sigma_i$ represents the uncertainty of the total backgrounds for the energy bin $i$. 
Here we take $\bar{v}_i$ as part of the observed event number per bin included in $n_i$.
We note that the total background uncertainty varies with DM mass since a different targeted $m_{\chi}$ leads to a different selection of the visible energy window for the analysis. 
The uncertainty is $19\%$, $16\%$, $20\%$, and $24\%$ for $m_{\chi}=15$, $20$, $30$, and \SI{40}{MeV}, respectively, while it is $25\%$ for $50\leq m_{\chi}/{\rm MeV}\leq 100$.
The number of energy bins is determined by the energy resolution $3\%/\sqrt{E[\rm{MeV}]}$ of JUNO. The degrees of freedom correspond to the number of energy bins. We note that ${\chi}^2$ equals to zero for $n_i$ = $v_i$ and ${v}_i$ = $\bar{v}_i$. Denoting $\chi^2$ in this case as $\chi^2_{\rm min}$, 
we then look for $\Delta \chi^2$ = $(\chi^2 - \chi^2_{\rm min})$ = $(1.645)^2$ to obtain the 90\% C.L. sensitivity limit for $\langle \sigma v\rangle$. The sensitivity limit depends on the DSNB flux we adopt as part of the total background. Table~\ref{table:DSNB_Para} summarizes four DSNB flux models adopted in this analysis. Hence, a specific sensitivity curve corresponds to a specific choice of DSNB flux as shown in Fig.~\ref{Upperlimit}.    
\paragraph{Bayesian analysis with Markov Chain Monte Carlo sampling}
The second approach is based on Bayesian analysis. A likelihood function is defined with the observed ($n_i$) and expected ($\lambda_i$) number of events in the $i$-th bin of the spectrum by 
\begin{equation}
    p(\textrm{spec}|S, B_{\rm DSNB}, ... , B_{\rm atmoNC}) = \prod_{i=1}^N \frac{\lambda_i(S, B_{\rm DSNB}, ... , B_{\rm atmoNC})^{n_i}}{n_i!}~e^{-\lambda_i(S, B_{\rm DSNB}, ... , B_{\rm atmoNC})}.
    \label{equation_likelihood}
\end{equation}
With Bayes' theorem, the posterior probability is $p(S,...,B_{\rm atmoNC}|\textrm{spec})\propto p(\textrm{spec}|S,..., B_{\rm atmoNC})\cdot p_0(S)\cdot\cdot\cdot p_0(B_{\rm atmoNC})$ with Eq.~(\ref{equation_likelihood}) and the prior probabilities for signal and background contributions $p_0$ where spec means the total spectrum. It represents the probability that the observed spectrum can be explained by the set of parameters $S$ and $B$ and is marginalized with respect to the background contributions
\begin{equation}
    p(S|\textrm{spec}) = \int p(S, B_{\rm DSNB},..., B_{\rm atmoNC}|\textrm{spec})~\textrm{d}B_{\rm DSNB}~\cdot\cdot\cdot \textrm{d}B_{\rm atmoNC}.
    \label{equation_marginalized_function}
\end{equation}
The 90\% probability upper limit $S_{90}$ on the number of signal events can therefore be calculated by equating the integral of Eq.~(\ref{equation_marginalized_function}) with a probability of 90\%: $\int_0^{S_{90}} p(S|\textrm{spec})~\textrm{d}S = 0.90$ \cite{Bayesian}. Data sets representing the expected number of events $\lambda_i$ are generated from the background-only spectrum following Poisson distribution and are analyzed for different mass settings of the DM signal. We set a flat prior probability of the signal contribution, $p_0(S) = 1/S_{\rm max}$ (with $S_{\rm max}=60$ according to the current limits of \cite{SuperK_782}). The prior probabilities of backgrounds are set by a Gaussian distribution with the mean value $\mu_B$ and width $\sigma_B$: 
\begin{equation}
p_0(B) = \textrm{exp}\left({-\frac{(B-\mu_B)^2}{2\sigma_B^2}}\right) \bigg/ {\displaystyle\int_0^\infty {\textrm{exp}\left({-\frac{(B-\mu_B)^2}{2\sigma_B^2}}\right) \textrm{d}B}},
\end{equation}
when $B \geq 0$. The marginalization of the posterior probability function is performed with Markov Chain Monte Carlo sampling using the Python package of \cite{EMCEE}, which is based on \cite{EMCEE_theory}. The 90\% upper limit $S_{90}$ of the number of signal events are calculated for each data set and the distribution of the values of $S_{90}$ can be interpreted as a probability density to determine the mean 90\% upper limit on the number of signal events as well as the upper limits on the anti-neutrino flux from DM self-annihilation and on the DM self-annihilation cross-section, respectively. 

Two different statistical approaches obtain consistent sensitivities to $\langle \sigma v\rangle$. In Fig.~\ref{Upperlimit}, we present JUNO's expected 90\% C.L. upper limit on $\langle \sigma v\rangle$ together with upper limits obtained or expected from other neutrino detectors. We analyze the total event spectrum for a $90.5\%$ PSD cut efficiency for IBD events and a $4.0\%$ PSD cut efficiency for Atm-$\nu$ NC background events with Bayesian analysis and a likelihood-ratio test. 
The spectral behaviors of the total background events and the DM signal could explain why the best search sensitivity occurs around $m_{\chi}=$\SI{40}{MeV}. Figs.~\ref{spectrum_a} and \ref{spectrum_b} show that the total number of background events gradually decreases with increasing visible energy until around \SI{40}{MeV}. On the other hand,  Fig.~\ref{figure_signal_spectrum} shows that the DM signal spectrum becomes broader with an increasing $m_{\chi}$. 
The best sensitivity of JUNO therefore occurs at $m_{\chi}\simeq$ \SI{40}{MeV}, which results in a $90\%$ upper limit on the thermally averaged DM annihilation rate, $\langle \sigma v\rangle(m_{\chi}=$ \SI{40}{MeV})=\SI{1.1e-25}{\cubic\centi\meter\per\second} for both Likelihood and Bayesian analysis over \SI{10}{years} of data taking.
The sensitivity curves in Fig.~\ref{Upperlimit}, except of those by JUNO, are scaled to the convention $J_{\rm avg}$ = 5. We compare our result with those of KamLAND~\cite{KamLAND}, SuperK~\cite{SuperK_782,SuperK_1542} and the HyperK expectation~\cite{HyperK}. We note that results by~\cite{SuperK_782} and~\cite{SuperK_1542} differ due to differences in the analyzed dataset and the background modeling as pointed out in~\cite{SuperK_1542}. JUNO will probe into $\langle \sigma v \rangle$'s that are up to an order of magnitude smaller than the ones SuperK obtained for the DM mass range of \SIrange[range-units = brackets, range-phrase=--]{30}{100}{MeV}. Such a sensitivity will be comparable to the one expected by HyperK in the DM mass range of \SIrange[range-units = brackets, range-phrase=--]{15}{100}{MeV}.
\begin{figure}
\centering
\includegraphics[width=0.95\textwidth]{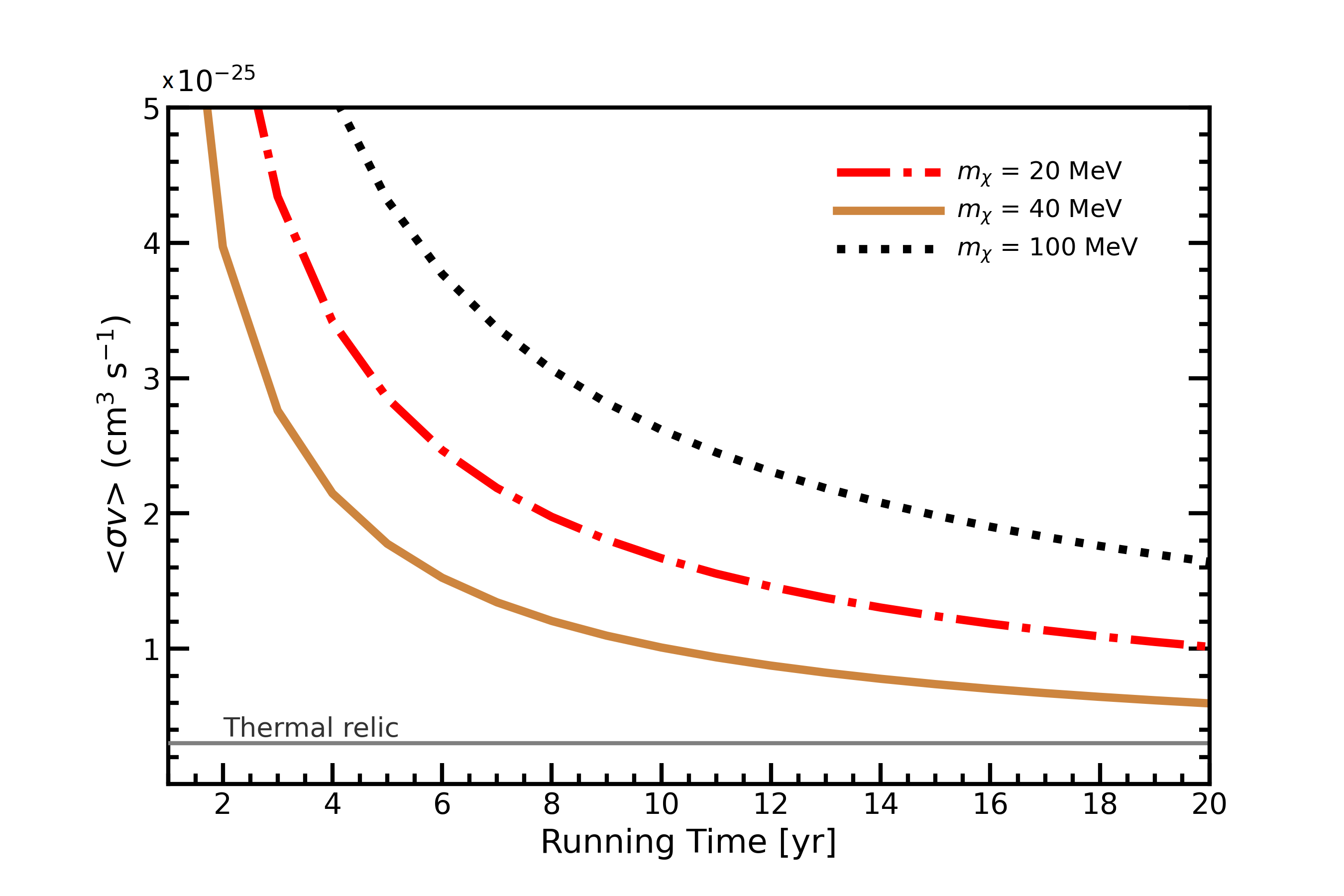}
\caption[90\% confidence level varies with time.]{The progress of JUNO 90\% C.L. sensitivities to $\langle \sigma v\rangle$ over the running time up to 20 years for DM masses of 20, 40, and \SI{100}{MeV}, respectively.  The thermal relic value $\langle \sigma v \rangle = 3 \times 10^{-26} \ \text{cm}^3\text{s}^{-1}$ is also shown for comparison. The nominal DSNB flux and the convention $J_{\text{avg}} = 5$ are adopted in the sensitivity calculations. }
\label{Upperlimit_Time}
\end{figure}

Besides presenting JUNO sensitivities to $\langle \sigma v\rangle$ for 10 years of data taking, we also present in Fig.~\ref{Upperlimit_Time} the progress of sensitivities over the running time for benchmark DM masses of 20, 40, and \SI{100}{MeV}, respectively. It is seen that, for $m_\chi=40$ MeV, the sensitivity to $\langle \sigma v\rangle$ can reach $6 \times 10^{-26} \ \text{cm}^3\text{s}^{-1}$ over 20 years of data taking.  
%-----------Section 7 Conclusion----------
\section{ Conclusion}
In this study, the JUNO sensitivity to the detection of neutrinos from DM self-annihilation in the Milky Way is investigated for the DM mass range of \SIrange[range-units = brackets, range-phrase=--]{15}{100}{MeV}. The expected $\bar{\nu}_e$ signal spectrum from the annihilation $\chi\chi \to \nu\bar{\nu}$ is evaluated with the JUNO offline simulation framework. Moreover, all possible background contributions to the above indirect DM search are investigated, which include the reactor $\bar{\nu}_e$ background, DSNB, atmospheric neutrino backgrounds from CC and NC interactions as well as muon-induced backgrounds. 
To further reduce the non-IBD Atm-$\nu$ NC and FN background events, a PSD method based on the tail-to-total ratio is applied. We apply a customized PSD cut and IBD signal selection based on each DM mass, which results in a different event spectrum and sensitivity from the DSNB study~\cite{JUNO_DSNB}.
The sensitivity of JUNO, i.e. the 90\% C.L. upper limit on the DM annihilation rate, $\langle \sigma v \rangle$, is analyzed based on both the Bayesian analysis and the Likelihood ratio test. JUNO could reach the sensitivity of $\langle \sigma v \rangle=$\SI{1.1e-25}{\cubic\centi\meter\per\second} with both Likelihood and Bayesian analysis for $m_{\chi}=$\SI{40}{MeV} over \SI{10}{years} of exposure. It corresponds to an improvement by a factor of 2 to 10 for the DM mass range of \SIrange[range-units = brackets, range-phrase=--]{15}{100}{MeV} compared to the present-day best limits obtained by SuperK and would be comparable to that expected by HyperK.
%----------- Acknowledgments ----------
\acknowledgments
We are grateful for the ongoing cooperation from the China General Nuclear Power Group. This work was supported by the Chinese Academy of Sciences, the National Key R\&D Program of China, the CAS Center for Excellence in Particle Physics, Wuyi University, and the Tsung-Dao Lee Institute of Shanghai Jiao Tong University in China, the Institut National de Physique Nucléaire et de Physique de Particules (IN2P3) in France, the Istituto Nazionale di Fisica Nucleare (INFN) in Italy, the Italian-Chinese collaborative research program MAECI-NSFC, the Fonds de la Recherche Scientifique (F.R.S-FNRS) and FWO under the “Excellence of Science - EOS” in Belgium, the Conselho Nacional de Desenvolvimento Cientifico e Tecnológico in Brazil, the Agencia Nacional de Investigacion y Desarrollo in Chile, the Charles University Research Centre and the Ministry of Education, Youth, and Sports in Czech Republic, the Deutsche Forschungsgemeinschaft (DFG), the Helmholtz Association, and the Cluster of Excellence PRISMA+ in Germany, the Joint Institute of Nuclear Research (JINR) and Lomonosov Moscow State University in Russia, the joint Russian Science Foundation (RSF) and National Natural Science Foundation of China (NSFC) research program, the NSTC and MOE in Taiwan, the Chulalongkorn University and Suranaree University of Technology in Thailand, and the University of California at Irvine in USA.
%-----------Reference----------
\bibliographystyle{unsrt}
\bibliography{main}

\providecommand{\noopsort}[1]{}\providecommand{\singleletter}[1]{#1}%
\begin{thebibliography}{10}

\bibitem{Rubin:1970}
V.C. Rubin and W.K. Ford.
\newblock {Rotation of the Andromeda Nebula from a Spectroscopic Survey of
  Emission Regions}.
\newblock {\em Astrophys. J.}, 159:379--403, 1970.

\bibitem{Jungman:1995df}
G.~Jungman, M.~Kamionkowski, and K.~Griest.
\newblock {Supersymmetric dark matter}.
\newblock {\em Phys. Rept.}, 267:195--373, 1996.

\bibitem{Super-Kamiokande:2004pou}
{The Super-Kamiokande Collaboration}, S.~Desai, et~al.
\newblock {Search for dark matter WIMPs using upward through-going muons in
  Super-Kamiokande}.
\newblock {\em Phys. Rev. D}, 70:083523, 2004.

\bibitem{IceCube_DMind}
{IceCube Collaboration}, R.~Abbasi, et~al.
\newblock {The design and performance of IceCube DeepCore}.
\newblock {\em Astropart. Phys.}, 35:615, 2012.

\bibitem{Fermi}
{The Fermi-LAT Collaboration}, M.~Ackermann, et~al.
\newblock {Searching for Dark Matter Annihilation from Milky Way Dwarf
  Spheroidal Galaxies with Six Years of Fermi Large Area Telescope Data}.
\newblock {\em Phys. Rev. Lett.}, 115:231301, 2015.

\bibitem{AMS2}
{AMS Collaboration}, M.~Aguilar, et~al.
\newblock {Observation of Complex Time Structures in the Cosmic-Ray Electron
  and Positron Fluxes with the Alpha Magnetic Spectrometer on the International
  Space Station}.
\newblock {\em Phys. Rev. Lett.}, 121:051102, 2018.

\bibitem{HESS}
{HESS Collaboration}, H.~Abdallah, et~al.
\newblock {Search for $\gamma$-Ray Line Signals from Dark Matter Annihilations
  in the Inner Galactic Halo from 10 Years of Observations with H.E.S.S.}
\newblock {\em Phys. Rev. Lett.}, 120:201101, 2018.

\bibitem{NuDM_MeV}
G.~Barenboim, P.B. Denton, and I.M. Oldengott.
\newblock {Constraints on inflation with an extended neutrino sector}.
\newblock {\em Phys. Rev. D}, 99:083515, 2019.

\bibitem{Accer}
C.A. Argüelles et~al.
\newblock {New opportunities at the next-generation neutrino experiments (Part
  1: BSM Neutrino Physics and Dark Matter)}.
\newblock {\em Rep. Prog. Phys.}, 83:124201, 2019.

\bibitem{Coiilders}
R.~Primulando and P.~Uttayarat.
\newblock {Dark Matter-Neutrino Interaction in Light of Collider and Neutrino
  Telescope Data}.
\newblock {\em JHEP}, 06:26, 2018.

\bibitem{SolarNu}
F.~Capozzi, I.M. Shoemaker, and L.~Vecchi.
\newblock {Solar Neutrinos as a Probe of Dark Matter-Neutrino Interactions}.
\newblock {\em JCAP}, 07:021, 2017.

\bibitem{KamLAND}
{KamLAND Collaboration}, A.~Gando, et~al.
\newblock {Search for extraterrestrial antineutrino sources with the KamLAND
  detector}.
\newblock {\em Astrophys. J.}, 745:193, 2012.

\bibitem{KamLAND_2022}
{KamLAND Collaboration}, S.~Abe, et~al.
\newblock {Limits on Astrophysical Antineutrinos with the KamLAND Experiment}.
\newblock {\em Astrophys. J.}, 925:14, 2022.

\bibitem{SKColab_DM}
{The Super-Kamiokande Collaboration}, K.~Abe, et~al.
\newblock {Indirect search for dark matter from the Galactic Center and halo
  with the Super-Kamiokande detector}.
\newblock {\em Phys. Rev. D}, 102:072002, 2020.

\bibitem{IceCUBEDM_Co}
{IceCube Collaboration}, R.~Abbasi, et~al.
\newblock {Search for GeV-scale dark matter annihilation in the Sun with
  IceCube DeepCore}.
\newblock {\em Phys. Rev. D}, 105:062004, 2022.

\bibitem{IceCUBE_DMEXP}
{ANTARES and IceCube Collaboration}, A.~Albert, et~al.
\newblock {Combined search for neutrinos from dark matter self-annihilation in
  the Galactic Center with ANTARES and IceCube}.
\newblock {\em Phys. Rev. D}, 102:082002, 2020.

\bibitem{SuperK_782}
A.~{Olivares-Del Campo}, C.~Boehm, S.~Palomares-Ruiz, and S.~Pascoli.
\newblock {Dark matter-neutrino interactions through the lens of their
  cosmological implications}.
\newblock {\em Phys. Rev. D}, 97:075039, 2018.

\bibitem{SuperK_1542}
C.A. Argüelles, A.~Diaz, A.~Kheirandish, A.~{Olivares-Del-Campo}, I.~Safa, and
  A.C. Vincent.
\newblock {Dark Matter Annihilation to Neutrinos}.
\newblock {\em Rev. Mod. Phys}, 92:035007, 2021.

\bibitem{Nu_Glalt_Icecube}
C.A. Argüelles, A.~Kheirandish, and A.C. Vincent.
\newblock Imaging galactic dark matter with high-energy cosmic neutrinos.
\newblock {\em Phys. Rev. Lett.}, 119:201801, 2017.

\bibitem{HyperK}
N.F. Bell, M.J. Dolan, and S.~Robles.
\newblock {Searching for sub-GeV Dark Matter in the Galactic Centre using
  Hyper-Kamiokande}.
\newblock {\em JCAP}, 09:019, 2020.

\bibitem{JUNO:2015zny}
{JUNO Collaboration}, F.~P. An, et~al.
\newblock {Neutrino Physics with JUNO}.
\newblock {\em J. Phys. G}, 43:030401, 2016.

\bibitem{Muon_strategies}
{JUNO Collaboration}, A.~Abusleme, et~al.
\newblock {JUNO physics and detector}.
\newblock {\em {Prog. Part. Nucl.}}, 123:103927, 2021.

\bibitem{Palomares-Ruiz:2007trf}
S.~Palomares-Ruiz and S.~Pascoli.
\newblock {Testing MeV dark matter with neutrino detectors}.
\newblock {\em Phys. Rev. D}, 77:025025, 2008.

\bibitem{Atmos_mu_e}
JUNO Collaboration, A.~Abusleme, et~al.
\newblock {JUNO sensitivity to low energy atmospheric neutrino spectra}.
\newblock {\em Eur. Phys. J. C}, 81:10, 2021.

\bibitem{JUNO_Calibration}
JUNO Collaboration, A.~Abusleme, et~al.
\newblock {Calibration strategy of the JUNO experiment}.
\newblock {\em JHEP}, 03:004, 2021.

\bibitem{JUNO_Top_Tracker}
JUNO Collaboration, A.~Abusleme, et~al.
\newblock {The JUNO experiment Top Tracker}.
\newblock {\em arXiv}, 2303.05172, 2023.

\bibitem{OPERA:2000zdf}
{OPERA Collaboration}, N.~Agafonova, et~al.
\newblock {Evidence for $\nu_\mu \to \nu_\tau$ appearance in the CNGS neutrino
  beam with the OPERA experiment}.
\newblock {\em Phys. Rev. D}, 89:051102, 2023.

\bibitem{OPERA_TOP_TRACKER}
T.~Adam et~al.
\newblock {The OPERA experiment target tracker}.
\newblock {\em Nucl. Instrum. Methods A}, 577:523, 2007.

\bibitem{NFW1996}
J.F. Navarro, C.S. Frenk, and S.D.M. White.
\newblock {The Structure of cold dark matter halos}.
\newblock {\em Astrophys. J.}, 462:563, 1996.

\bibitem{KKBP1998}
A.V. Kravtsov, A.A. Klypin, J.S. Bullock, and J.R. Primack.
\newblock {The Cores of dark matter-dominated galaxies}.
\newblock {\em Astrophys. J.}, 502:48, 1998.

\bibitem{MQGSL1999}
B.~Moore, T.~Quinn, F.~Governato, J.~Stadel, and G.~Lake.
\newblock {Cold collapse and the core catastrophe}.
\newblock {\em Mon. Not. R. Astron. Soc.}, 310:1147, 1999.

\bibitem{IBDcross}
A.~Strumia and F.~Vissani.
\newblock {Precise quasielastic neutrino/nucleon cross section}.
\newblock {\em Phys. Lett. B}, 564:42, 2003.

\bibitem{IBDangle}
P.~Vogel and J.~F. Beacom.
\newblock {Angular distribution of neutron inverse beta decay, $\bar{\nu_e}$ +
  p $\rightarrow$ $e^+$ + n}.
\newblock {\em Phys. Rev. D}, 60:053003, 1999.

\bibitem{Mueller}
T.~A. Mueller et~al.
\newblock {Improved Predictions of Reactor Antineutrino Spectra}.
\newblock {\em Phys. Rev. C}, 83:054615, 2011.

\bibitem{Huber}
P.~Huber.
\newblock {On the determination of anti-neutrino spectra from nuclear
  reactors}.
\newblock {\em Phys. Rev. C}, 84:024617, 2011.

\bibitem{DSNBflux}
S.~Ando and K.~Sato.
\newblock {Relic neutrino background from cosmological supernovae}.
\newblock {\em New J. Phys}, 6:179, 2004.

\bibitem{RSN}
S.~Horiuchi, J.~F. Beacom, and E.~Dwek.
\newblock {The Diffuse Supernova Neutrino Background is detectable in
  Super-Kamiokande}.
\newblock {\em Phys. Rev. D}, 79:083013, 2009.

\bibitem{E_avg_BHR}
A.~Priya and C.~Lunardini.
\newblock {Diffuse neutrinos from luminous and dark supernovae: prospects for
  upcoming detectors at the $O$(10) kt scale}.
\newblock {\em JCAP}, 11:031, 2017.

\bibitem{DSNB_upperflux_SK}
{The Super-Kamiokande Collaboration}, K.~Abe, et~al.
\newblock {Diffuse supernova neutrino background search at Super-Kamiokande}.
\newblock {\em {Phys. Rev. D}}, 104:122002, 2021.

\bibitem{Honda}
M.~Honda, M.~Sajjad Athar, T.~Kajita, K.~Kasahara, and S.~Midorikawa.
\newblock {Atmospheric neutrino flux calculation using the NRLMSISE-00
  atmospheric model}.
\newblock {\em {Phys. Rev. D}}, 92:023004, 2015.

\bibitem{Battistoni:2015epi}
G.~Battistoni et~al.
\newblock {Overview of the FLUKA code}.
\newblock {\em {Annals Nucl. Energy}}, 82:10, 2015.

\bibitem{GENIE}
C.Andreopoulos et~al.
\newblock {The GENIE Neutrino MC Generator}.
\newblock {\em {Nucl. Instrum. Meth. A}}, 87:614, 2010.

\bibitem{12C_Xsec}
T.~Yoshida et~al.
\newblock {Neutrino-Nucleus Reaction Cross Sections for Light Element Synthesis
  in Supernova Explosions}.
\newblock {\em {Astrophys. J.}}, 686:448, 2008.

\bibitem{Isotope_KamLAND}
{KamLAND Collaboration}, S.~Abe, et~al.
\newblock {Production of Radioactive Isotopes through Cosmic Muon Spallation in
  KamLAND}.
\newblock {\em {Astrophys. J.}}, 81:025807, 2010.

\bibitem{Isotope_Borexino}
{Borexino Collaboration}, G.~Bellini, et~al.
\newblock {Cosmogenic Backgrounds in Borexino at 3800m water-equivalent depth}.
\newblock {\em {JCAP}}, 049:1308, 2013.

\bibitem{NC_12C_AtmNu-I}
J.~Cheng, Y.-F. Li, H.-Q. Lu, and L.-J. Wen.
\newblock {Neutral-current background induced by atmospheric neutrinos at large
  liquid-scintillator detectors. I. Model predictions}.
\newblock {\em {Phys. Rev. D}}, 103:053001, 2021.

\bibitem{GENIE_FSI}
S.~Dytman.
\newblock {Final state interactions in neutrino–nucleus experiments}.
\newblock {\em {Acta Phys. Pol. B}}, 40:2445, 2009.

\bibitem{TALYS}
A.~Koning, S.~Hilaire, and M.~Duijvestijn.
\newblock {TALYS: Comprehensive Nuclear Reaction Modeling}.
\newblock {\em AIP Conf. Proc}, 769:1154, 2005.

\bibitem{12C_deexc_model1}
Y.~A. Kamyshkov and E.~Kolbe.
\newblock {Signatures of nucleon disappearance in large underground detectors}.
\newblock {\em {Phys. Rev. D}}, 67:076007, 2003.

\bibitem{12C_deexc_model2}
N.~Auerbach, N.~Van Giai, and O.~Vorov.
\newblock {Neutrino scattering from C-12 and O-16}.
\newblock {\em {Phys. Rev. C}}, 56:2368, 1997.

\bibitem{12C_deexc_model3}
E.~Kolbe, K.~Langanke, and P.~Voge.
\newblock {Weak reactions on C-12 within the continuum random phase
  approximation with partial occupancies}.
\newblock {\em {Nucl. Phys. A}}, 652:91, 1999.

\bibitem{NC_12C_AtmNu-II}
J.~Cheng, Y.-F. Li, H.-Q. Lu, and L.-J. Wen.
\newblock {Neutral-current background induced by atmospheric neutrinos at large
  liquid-scintillator detectors. II. Methodology for in situ measurements}.
\newblock {\em {Phys. Rev. D}}, 103:053002, 2021.

\bibitem{LENA_T2T}
R.~Mollenberg et~al.
\newblock {Detecting the Diffuse Supernova Neutrino Background with LENA}.
\newblock {\em Phys. Rev. D}, 91:032005, 2015.

\bibitem{TMVA}
A.~Hocker et~al.
\newblock {TMVA - Toolkit for Multivariate Data Analysis}.
\newblock {\em ArXiv}, 0703039, 2009.

\bibitem{Skitlearn}
F.~Pedregosa et~al.
\newblock Scikit-learn: Machine learning in python.
\newblock {\em J. Machine Learning Res}, 12:2825, 2011.

\bibitem{JUNO_DSNB}
{JUNO Collaboration}, A.~Abusleme, et~al.
\newblock {Prospects for detecting the diffuse supernova neutrino background
  with JUNO}.
\newblock {\em {JCAP}}, 10:033, 2022.

\bibitem{Bayesian}
A.~Caldwell and K.~Kröninger.
\newblock {Signal discovery in sparse spectra: A Bayesian analysis}.
\newblock {\em Phys. Rev. D}, 74:092003, 2006.

\bibitem{EMCEE}
D.~Foreman-Mackey, D.W. Hogg, D.~Lang, and J.~Goodman.
\newblock {emcee: The MCMC Hammer}.
\newblock {\em Publ. Astron. Soc. Pac.}, 125:306, 2013.

\bibitem{EMCEE_theory}
J.~Goodman and J.~Weare.
\newblock Ensemble samplers with affine invariance.
\newblock {\em Commun. Appl. Math. Comput. Sci.}, 5(1):65, 2010.

\end{thebibliography}
\end{document}